%% file: main.tex
\documentclass[11pt]{article}
\usepackage{natbib}
\usepackage{diagbox}
\usepackage{amsmath,bm}
\usepackage{float}
\usepackage{xcolor}
\usepackage{url}
\usepackage{subcaption}
\usepackage{booktabs}
\usepackage{soul}
\usepackage{authblk}
\usepackage[symbol]{footmisc}

\input{structure.tex} 

\title{A Latent Trajectory Analysis for Multivariate Outcomes with Mixed-Scale: Application to Alzheimer’s Disease Neuroimaging Initiative}
\date{}

\author[1]{Lindsay R. Salvati}
\author[2]{Jungwun Lee\footnote{Corresponding author}}

\affil[1]{Department of Biostatistics, Boston University School of Public Health, Boston, MA, USA}
\affil[2]{Department of Biostatistics, Boston University School of Public Health, Boston, MA, USA}

\begin{document}
\maketitle 

\section{Introduction}\label{intro}
\noindent Alzheimer's disease (AD) is a progressive neurodegenerative disorder characterized by memory loss, cognitive decline, and functional impairment. Currently, an estimated 6.9 million Americans aged 65 or older live with Alzheimer's dementia \citep{better2023alzheimer}. Despite this massive global prevalence and the profound burden on individuals and communities, there remains no definitive cure. Consequently, research priorities have shifted toward identifying effective strategies to prevent and slow the progression of AD \citep{bateman_two_2023}. Large-scale initiatives, such as the Alzheimer's Disease Neuroimaging Initiative (ADNI), have been established to collect comprehensive longitudinal clinical, imaging, genetic, and biomarker data. These datasets contain a wealth of mixed-type variables: continuous measures (e.g., cognitive test scores, BMI, biomarkers) and categorical items (e.g., diagnostic status, presence of depression). Analyzing these variables concurrently allows for the investigation of disease evolution (or trajectories) across the entire study sample or within pre-specified subgroups \citep{nguena_nguefack_trajectory_2020}.  

\noindent Unfortunately, modeling the latent trajectories of multivariate continuous and categorical variables simultaneously presents a significant statistical challenge. Analyses must account for the associations between variable types at specific time points while simultaneously handling the correlations arising from repeated measurements. Traditional approaches, such as mixed-effects models, typically assume a single population mean trajectory, treating individual variation as random effects around that mean. This assumption may obscure meaningful heterogeneity in disease progression. Specifically in AD, clinical manifestations of those with the disease can often present very differently, especially depending on the stage of progression. Some patients remain stable, while others decline rapidly. Furthermore, decline may manifest across different functional domains (e.g., memory vs. language) or be compounded by comorbidities. To address this, researchers utilize latent trajectory modeling, which assumes the population is composed of distinct subgroups (latent classes). Individuals within a class share common longitudinal characteristics, while those in different classes exhibit heterogeneity. Identifying these latent subgroups can provide critical insights into differential risk profiles and treatment responses. For example, identifying a subgroup of patients likely to show rapid progression could allow clinical trialists to select a more homogeneous cohort, thereby increasing statistical power and the likelihood of detecting treatment effects.


\noindent Growth mixture modeling (GMM) \citep{jung_introduction_2008, muthen_integrating_2000,muthen_chapter_2004} is a standard way for identifying latent classes of population based on longitudinal behaviors. However, standard GMM frameworks have notable limitations. For example, the traditional growth mixture modeling framework \citep{muthen_chapter_2004, jung_introduction_2008} can only model a single outcome variable and thus cannot be employed for a longitudinal study with multivariate outcomes. Latent growth mixture model \citep{proust-lima_analysis_2013, proust-lima_estimation_2017, koo2020bayesian, proust-lima_describing_2023} and dynamic structural model \citep{proust-lima_describing_2023-1, tadde_dynamic_2020} are capable of multivariate outcome variables, but their identification of latent trajectories relies on the marginal mean trends of outcome variables via parametric form of mean vectors and variance-covariance matrices. Such parametric models require correct specifications on the trajectory-specific mean and variance-covariance structures, which may seriously mislead conclusions if incorrectly specified. Also, a single-layer approach captures only longitudinal response trends and cannot identify response patterns at a particular time period, which is often of interest.

\noindent As an alternative to parametric GMMs, Latent Class Profile Models (LCPM) \citep{chung_latent_2011, lee_bayesian_2021, lee_latent_2024} and Latent Transition Models (LTM) \citep{chung_latent_2008} offer a non-parametric approach to capturing longitudinal patterns. Rooted in Latent Class Model (LCM) \citep{lazarsfeld_latent_1968, goodman_exploratory_1974}, these methods treat the trajectory shapes flexibly without imposing strict functional forms. While standard LCMs were designed for cross-sectional categorical data, the LCPM extends this framework to longitudinal settings by identifying (1) latent classes at each time point and (2) representative sequential transition patterns of class membership over time. Similarly, for a set of repeatedly measured categorical response variables, LTM identifies (1) latent classes based on the response variables at the initial time point, and (2) transition matrices between two adjacent time point that describe how individuals move across latent classes as time flows. Specifically, these latent models identify the subgroups of populations (i.e., latent classes) based on individuals’ outcome patterns at baseline and describe how individual-level latent class memberships change over time. While these non-parametric approaches can be robust to the misspecification of the functional form and can be much more interpretable, they are only applicable for categorical outcomes, meaning they cannot currently accommodate continuous measurements, such as biomarkers or cognitive scores, which are essential for fully capturing the nuance of AD progression. To address this gap, we propose extending the existing LCPM framework to model joint trajectories of both continuous and categorical outcomes. 

\noindent This paper aims to address the following research questions: (1) What are the representative patterns of multivariate mixed outcomes and how do they interplay at each time point? (2) What are the representative latent trajectory patterns of these outcomes over time? (3) How are individual-level covariates associated with latent trajectory membership? and (4) How can we appropriately handle missing values in multivariate longitudinal outcomes? We demonstrate the utility of this method through an application to the ADNI database. The remainder of the paper is organized as follows: Section 2 details the methods, including the conventional LCPM, our proposed extension, and the maximum-likelihood (ML) estimation strategy using the expectation-maximization (EM) algorithm \citep{dempster_maximum_1977}. Section 3 describes the simulation study, followed by the ADNI data application in Section 4. We conclude with a discussion of implications and future directions.

\section{Methods}\label{Methods}

\subsection{Latent Class Profile Model: A Review}\label{LCPM}

The latent class profile model (LCPM) was proposed by Chung et al. \citep{chung_latent_2011}. Let $\textbf{Y}_{t}=[Y_{1t},\dots,Y_{Mt}]$ be a vector of $M$ categorical variables, and $\textbf{y}_{it}=[y_{i1t},\dots,y_{iMt}]$ be the realized outcomes for the $i$th subject at time $t$, $t=1,\dots,T$, where each $y_{imt}$ can have exactly one of $1,\dots,r_m$ possible outcomes. Two layers of latent variables describe the overall trajectories of subjects. The first layer is a vector of categorical latent variables $\textbf{C}= [C_1, \dots, C_T]$, where each $C_t$ has $K$ categories and each category represents a particular response pattern of $\textbf{Y}_{t}$ at time $t$. While $\textbf{C}$ follows a multinomial distribution with $K^T$ categories, not every category is equally important; such a large number of patterns can be efficiently summarized into $S$ categories, which constitutes a latent profile variable $U$. This latent profile variable $U$ summarizes the joint distribution of $\textbf{C}$ into $S$ representative categories, by introducing another conditional independence assumption among $\textbf{C}$; such that $[C_1, \dots, C_T]$ are conditionally independent given a latent profile membership $U = u$ so that subjects with the same latent profile membership are homogeneous in their latent class memberships. Finally, associations between subject-specific covariates $\textbf{x}_i=[x_{i1},x_{i2},\dots, x_{iP}]$ and latent profile memberships can be specified via a multinomial regression. In this way, the joint distribution of longitudinal outcomes ${\bf Y} = [{\bf Y}_{1},\ldots,{\bf Y}_{T}]$ can be decomposed as follows:
\begin{equation}\label{LCPA-original}
\begin{aligned}
        P(\textbf{Y}\mid \textbf{x}) &= \sum\limits_{u=1}^S \sum\limits_{c_1=1}^K \dots \sum\limits_{c_T=1}^K P(\textbf{Y},\textbf{C}=\textbf{c},U=u\mid\textbf{x}) \\
       &= \sum\limits_{u=1}^S \sum\limits_{c_1=1}^K \dots \sum\limits_{c_T=1}^K P(U=u\mid\textbf{x})P(\textbf{C}=\textbf{c}\mid U=u)P(\textbf{Y}=\textbf{y}\mid\textbf{C}=\textbf{c}) \\
       &=\sum\limits_{u=1}^S \sum\limits_{c_1=1}^K \dots \sum\limits_{c_T=1}^K P(U=u\mid\textbf{x}) \prod\limits_{t=1}^T \left[ P(C_t=c_t\mid U=u)\prod\limits_{m=1}^M P(Y_{mt}\mid C_t = c_t)\right].
\end{aligned}
\end{equation}

\noindent Since all outcomes and latent variables are categorical, all conditional probabilities can be described using multinomial distributions. \citep{chung_latent_2011} Specifically, the observed data likelihood of Eq.(\ref{LCPA-original}) can be written as follows:

\begin{equation}  \label{LCPA-original2}
\begin{aligned}
        P(\textbf{Y}\mid \textbf{x}) &=\sum\limits_{u=1}^S \sum\limits_{c_1=1}^K \dots \sum\limits_{c_T=1}^K P(U=u\mid\textbf{x}) \prod\limits_{t=1}^T \left[ P(C_t=c_t\mid U=u)\prod\limits_{m=1}^M P(Y_{mt}\mid c_t)\right] \\
       &=\sum\limits_{u=1}^S \sum\limits_{c_1=1}^K \dots \sum\limits_{c_T=1}^K \gamma_u(\textbf{x}) \prod\limits_{t=1}^T \left[ \eta_{c_t|u} ^{(t)} \prod\limits_{m=1}^M \prod\limits_{k=1}^{r_m} \rho_{mk|c_t}^{I(y_{imt}=k)} \right],
\end{aligned}
\end{equation}

\noindent where the parameters and assumptions have been specified according to Chung et al. \citep{chung_latent_2011} They can be interpreted as the following:

\begin{enumerate}
    \item $\rho_{mk|c_t}=P(Y_{mt}=k|c_t)$ represents the probability of a response variable $y_{mt}$ having value $k$ given latent class membership $c_t$ at time $t$. 
    \item $\eta_{c_t|u} ^{(t)}=P(C_t=c_t|U=u)$ represents the probability of being in class $c_t$ at time $t$ given latent profile membership $u$. 
    \item $\gamma_u=P(U=u)$ represents the probability of belonging to latent profile $u$. This can be a function of time-indepedent covariates $\textbf{x}$, where
    \begin{align*}
        \gamma_u(\textbf{x})=P(U=u|\textbf{x})= \frac{\text{exp}(\textbf{x}\beta_u)}{\sum\limits_{s=1}^S\text{exp}(\textbf{x}\beta_s)},~~\beta_u \in \mathbb{R}^P, \beta_1=0,u=1,\dots,S.
    \end{align*}
\end{enumerate}

\subsection{Mixed Latent Class Profile Model (Mixed-LCPM)}\label{MixedLCPM}

We propose a natural extension of the original LCPM by expanding the distribution of categorical outcomes to mixed-scale outcomes. Suppose we have both a vector of $P_1$ continuous outcomes $\textbf{Y}_{t}=[Y_{1t},\ldots,Y_{P_1t}]$, and a vector of $P_2$ categorical outcomes $\textbf{Z}_{t}=[Z_{1t},\dots,Z_{P_2t}]$ measured at each time point $t$, $t = 1,\ldots, T$. We denote $[\textbf{Y},\textbf{Z}]$ as the collection of all outcomes $[\textbf{Y}_{1},\textbf{Z}_{1},\ldots,\textbf{Y}_{T},\textbf{Z}_{T}]$ measured across $T$ time periods. Next, the proposed model follows identically from the standard LCPM; we posit a vector of categorical latent variables $\textbf{C}= [C_1, \ldots, C_T]$ that represent subject-level response patterns at time $t$, and a latent profile variable $U$ that summarizes the patterns of $\textbf{C}$ into $S$ representative categories. We impose three conditional independence assumptions: (1) continuous and categorical outcomes $[\textbf{Y}_t, \textbf{Z}_t]$ are conditionally independent give a latent class membership $c_t$ measured at the same time $t$, (2) categorical outcomes $\textbf{Z}_{t}=[Z_{1t},\ldots,Z_{P_2t}]$ are conditionally independent given $C_t$, (3) latent class variables $\textbf{C}= [C_1,\ldots,C_T]$ are conditionally independent given a latent profile membership $U = u$. Based on such assumptions, the joint distribution of outcomes $[\textbf{Y},\textbf{Z}]$ given covariates $\textbf{x}$ can be decomposed as follows: 
\begin{equation}  \label{LCPA-mixed}
\begin{aligned}
       P(\textbf{Y},\textbf{Z}\mid\textbf{x}) &= \sum\limits_{u=1}^S \sum\limits_{c_1=1}^K \dots \sum\limits_{c_T=1}^K P(\textbf{Y},\textbf{Z},\textbf{C}=\textbf{c},U=u\mid\textbf{x}) \\
       &= \sum\limits_{u=1}^S \sum\limits_{c_1=1}^K \dots \sum\limits_{c_T=1}^K P(U=u|\textbf{x})P(\textbf{C}=\textbf{c}\mid U=u)P(\textbf{Y}\mid\textbf{C}=\textbf{c})P(\textbf{Z}\mid\textbf{C}=\textbf{c}) \\
       &=\sum\limits_{u=1}^S \sum\limits_{c_1=1}^K \dots \sum\limits_{c_T=1}^K P(U=u\mid\textbf{x}) \prod\limits_{t=1}^T \left[ P(C_t=c_t\mid U=u)P(\textbf{Y}\mid C_t=c_t)\prod\limits_{p=1}^{P_2}P(Z_{pt}\mid C_t=c_t)\right]. 
\end{aligned}
\end{equation}

\noindent Similar to the standard LCPM, the proposed model assumes that (1) associations between outcome variables at the same time point are fully explained by the latent class variable at the corresponding time point, and (2) associations between outcome variables from different points are fully explained by the latent profile variable only through latent class memberships. Note that the conditional independence assumption on outcomes only applies to categorical outcomes; categorical outcomes $\textbf{Z}_{t}=[Z_{1t},\dots,Z_{P_2t}]$ are conditionally independent given latent class membership $c_t$, and they are also independent of continuous outcomes $\textbf{Y}_{t}$. Such conditional independence assumptions are key components of describing the joint distribution of multivariate outcomes with mixed scales. We also introduce several distributional assumptions to derive the likelihood function of the proposed model. First, we assume that continuous outcomes follow a multivariate normal distribution with class-specific means and covariance matrices: $\textbf{Y} \sim \textit{MVN}_{P_1}(\mu_c, \Sigma_c)$. Next, categorical outcomes and latent variables naturally follow multinomial distributions: $[Z_{pt}| C_t=c] \sim  \textit{multinom}(1,\pi_{p1|c},\dots,\pi_{pr_p|c})$ where $Z_{pt}$ has $r_p$ categories, $p = 1,\dots,P_2,$ and $t=1,\dots,T$. Parameters associated with these distributional assumptions are defined as follows:

\begin{itemize}
    \item $\pi_{pk\mid c}=P(Z_{ipt}=k\mid c)$ represents the probability of the categorical variable $z_{pt}$ having value $k$ given latent class membership $c$ at time $t$. 
    \item ${I(z=k)}$ is the indicator function which takes value 1 if $z$ is equal to $k$ and 0 otherwise
    \item $\phi(\textbf{Y}\mid \mu, \Sigma)$ is a function of the multivariate normal distribution such that $\textbf{Y} \sim \textit{MVN}_{P_1}(\mu_c, \Sigma_c)$
    \item $\eta_{c_t\mid u}^{(t)}=P(C_t=c_t\mid U=u)$ represents the probability of being in class $c_t$ at time $t$ given latent profile membership $u$. 
    \item $\gamma_u=P(U=u)$ represents the probability of belonging to latent profile $u$.
\end{itemize}

\noindent The $\pi$-parameter, which we denote as \textit{primary measurement parameter}, describes how individuals in each class respond to the $p$th categorical item at each time point. The $\mu$-parameter is also a \textit{primary measurement parameter}, and this describes the mean vector for the continuous responses of individuals in each class. We refer to the $\eta$-parameter as a \textit{secondary measurement parameter}. This describes the relationship between class $c_t$ at time $t$ and a class profile $u$. We identify a class profile by the set of estimated secondary measurement parameters from an individual’s sequential pattern of class membership. Note that measurement parameters ($\pi, \mu, \Sigma$) are time-invariant to ensure the interpretations on estimated latent classes are consistent across all time periods. This not only reduces the complexity of the model, but also ensures that a specific latent class (e.g., "High Risk") retains the same definition at $t=1$ as it does at $t=T$. Such a time-invariant assumption may be seemingly unrealistic, but it allows time-dependent associations between outcomes to be captured by the second-level latent structure (i.e., latent class variables and the latent profile variable), so that the temporal correlations and longitudinal trends are purely expressed via longitudinal transitions of latent class memberships. Further, the time-invariant assumption can be tested using the likelihood-ratio test. Finally, the prevalence parameter $\gamma$ can also be replaced with a logistic function of subject-level covariates $\textbf{x}_i$ as follows:

    \begin{align} \label{beta}
   \gamma_u(\textbf{x}_i)=P(U=u\mid\textbf{x}_i)= \frac{\text{exp}(\textbf{x}_i'\beta_u)}{\sum\limits_{s=1}^S\text{exp}(\textbf{x}_i'\beta_s)},~~\beta_u \in \mathbb{R}^P, \beta_1=0,u=1,\dots,S.
    \end{align}
    
\noindent In Equation \ref{beta}, $\beta_u$ represents associations between demographic and/or time-invariant factors $\textbf{x}_i$ and the $u$th  latent profile using multinomial logistic regression. In such a way, the proposed model describes associations between the latent profile variable membership $U = u$ and subject-specific covariates $\textbf{x}_i=[x_{i1},x_{i2},\dots, x_{iP}]$ in odd-ratio scales.

\noindent Based on the parameters described above and from the joint distribution in Equation \ref{LCPA-mixed}, we can explicitly calculate the complete-data likelihood of subject $i$, which is equivalent to the joint probability that $i$th individual belongs to the sequence $c$ and the class profile $u$ and provides responses $\textbf{y}_{it}=[y_{i1t},\dots,y_{iP_1t}]$ and $\textbf{z}_{it}=[z_{i1t},\dots,z_{iP_2t}]$, given the individual’s sequence of class membership over time, $\textbf{c}= [c_1,\dots,c_T]$, as follows: 

\begin{equation}  \label{completedatalikelihood}
\begin{aligned}
L^\star(u, \textbf{c}\mid\textbf{y}_i, \textbf{z}_i, \textbf{x}_i) &= P(\textbf{Y}_i=\textbf{y}_i,\textbf{Z}_i=\textbf{z}_i,\textbf{C}=\textbf{c},U=u\mid\textbf{x}_i) \\ 
&= \gamma_u(\textbf{x}_i) \prod\limits_{t=1}^T \left[ \eta_{c_t\mid u} ^{(t)} \phi(\textbf{y}_{it}\mid \mu_{c_t,t}, \Sigma_{c_t,t}) 
       \prod\limits_{p=1}^{P_2} \prod\limits_{k=1}^{r_p} \pi_{pk\mid c_t}^{I(z_{ipt}=k)} \right].
\end{aligned}
\end{equation}

\noindent Finally, the observed data likelihood of subject $i$, denoted as $L(\textbf{y}_i, \textbf{z}_i, \textbf{x}_i)$, can be obtained by integrating the complete data likelihood over all possible values of latent variables as follows: 

\begin{equation}  \label{obsdatalikelihood}
\begin{aligned}
L(\textbf{y}_i, \textbf{z}_i, \textbf{x}_i) &= \sum\limits_{u=1}^S \sum\limits_{c_1=1}^K \dots \sum\limits_{c_T=1}^K \gamma_u(\textbf{x}_i) \prod\limits_{t=1}^T \left[ \eta_{c_t\mid u} ^{(t)} \phi(\textbf{y}_{it}\mid\mu_{c_t,t}, \Sigma_{c_t,t}) 
       \prod\limits_{p=1}^{P_2} \prod\limits_{k=1}^{r_p} \pi_{pk\mid c_t}^{I(z_{ipt}=k)} \right].   
\end{aligned}
\end{equation}

\noindent Such a two-level structure offers two advantages over traditional single-level latent models. First, it separates the response patterns and trajectory patterns into two latent structures. It thus identifies both (1) latent classes of response patterns at each time and (2) transitional trends of latent class memberships simultaneously. In this way, the model's interpretability improves. Second, it allows greater flexibility in characterizing longitudinal trends in multivariate outcomes, as it does not assume a parametric form when describing trajectories. Instead, it describes trajectories in a nonparametric manner; trajectories are characterized by their sequential patterns of latent class memberships and thus do not require parametric assumptions about trajectories.

\subsection{Estimation and Inference} \label{estimation} 

This section discusses parameter estimation strategy for the proposed Mixed-LCPM via maximum likelihood (ML) estimation. Finding the ML estimates of the proposed model can be considered as a missing data problem, since the observed data likelihood of the proposed model is related with the latent class and profile memberships that cannot be directly observed from the data. As a result, we propose using the ML estimates using the EM algorithm \citep{dempster_maximum_1977}, an iterative technique for finding ML estimates from incomplete data.
The EM algorithm iteratively alternates between calculating the expected value of the complete data log-likelihood with respect to the posterior distribution of the latent variables (E-step) and updating the parameters to maximize this expectation (M-step). 

\noindent The \textbf{Expectation (E) step} calculation involves the full conditional probability of profile membership $u$ and class sequence $\textbf{c}$ given the observed data $(\textbf{y}_i, \textbf{z}_i)$ and covariates $\textbf{x}_i$. Direct calculation of the full posterior $\theta_{i(u,\textbf{c})}$ is computationally expensive due to the dimensionality of the class sequences \citep{lee_latent_2025}. The dimension increases exponentially for $T$, ($K^T \times S$), and so for large T this becomes non-negligible. To address this, we utilize the forward-backward algorithm (a recursive formula) \citep{bartolucci_overview_2010} to efficiently decompose the joint probability into a forward probability $P(Y_1=y_1,\ldots,Y_t=y_t, C_t=c_t \mid u)$ and a backward probability $P(Y_{t+1}=y_{t+1},\ldots,Y_T=y_T \mid U=u)$. 

\noindent In the \textbf{Maximization (M) step}, we maximize the expected complete data log-likelihood with respect to the model parameters using a Lagrange multiplier. Under the assumption that continuous outcomes follow a multivariate normal distribution given the latent class, closed-form solutions exist for the expectations. We include an additional step using the Newton-Raphson algorithm, since closed forms for $\beta$ cannot be expressed, to account for this. Lastly, under the missingness at random (MAR) assumption, we can also extend the algorithm to account for missing observations on outcome variables. Full derivations and algorithmic details are provided in the Supplementary Material.

\noindent It is known that a solution from the EM algorithm is not necessarily the global maximizer of the observed data likelihood. To avoid such a local maxima problem, one must try several initial values and choose the one with the highest likelihood \citep{wu_convergence_1983,dempster_maximum_1977}. Also, prior to the analysis, the number of latent classes $K$ and profiles $S$ must be specified, which cannot be observed from the data. As a result, we determine the optimal model structure using the Bayesian Information Criterion (BIC) \citep{schwarz_estimating_1978}, where a smaller BIC is preferred. Following Bandeen-roche et al. \citep{bandeen-roche_latent_1997}, model selection is performed without covariates, based off the marginalization property. 

\noindent To calculate standard errors of the EM estimates, we calculate the second derivative of the logarithm of the observed data likelihood (i.e., Hessian matrix), evaluate it at the MLE, and take the negative of its inverse. Then the square root of the diagonal elements of the information matrix becomes the asymptotic standard error of the EM estimates. Full derivations and details of the Hessian matrix of the proposed model are provided in the Supplementary Material. Throughout the simulation studies, we demonstrate that these asymptotic standard errors yield proper confidence intervals with valid coverage probabilities.

\noindent Our proposed estimation stems from the traditional EM algorithm and thus enjoys all theoretical and numerical properties of ML estimates given certain conditions are met \citep{wu_convergence_1983}. Let $\theta = [\boldsymbol{\mu}, \boldsymbol{\Sigma}, \boldsymbol{\rho}, \boldsymbol{\eta}, \boldsymbol{\gamma}]$ be the vector of all parameters of Mixed-LCPM, and let $\Omega$ be the parameter space. We denote $Q(\theta^{'}\mid \theta) = E[\log L^\star(u, \textbf{c} \mid \textbf{y}_i, \textbf{z}_i, \textbf{x}_i)]$ as the expected value of log-complete data likelihood evaluated at $\theta^{'} \in \Omega$, given observed data $[\textbf{y}_i, \textbf{z}_i, \textbf{x}_i]$ and current parameter value $\theta$. Specifically, our EM estimates achieves convergence to one of the local maxima if following conditions are satisfied \citep{wu_convergence_1983}:

\begin{enumerate}
    \item The expected value of the log-complete data likelihood, $Q(\theta^{*}\mid \theta)$, is continuous in $\Omega$, including the current value of the parameter $\theta$, and the updated parameter value $\theta^{*}$.
    \item $\text{sup}_{\theta^{*}\in \Omega}Q(\theta^{*}\mid \theta) > Q(\theta\mid \theta)$ for all $\theta^{*},\theta \in \Omega \backslash \Omega^{'}$, where $\Omega^{'}$ denotes the set of stationary points in the interior of $\Omega$.
\end{enumerate}

\noindent Although the first condition cannot be explicitly verified, we claim that it is satisfied because our distributional assumptions lie within the exponential family. Also, the standard EM algorithm is known to satisfy the second condition\citep{wu_convergence_1983}. Further, in Section \ref{simulation}, we illustrate that the proposed EM algorithm yields approximately unbiased point estimates and $95\%$ confidence intervals with nominal coverage probabilities. Such results indicate that our proposed estimation is valid given an appropriate initial value(that does not fall into one of the stationary points) and correctly specified numbers of latent classes and profiles. To assure the algorithm yields the global maximum, it is essential to (1) repeat a model fitting using a large number of initial values and choose the result with the highest likelihood value, and (2) repeat step (1) several times to check if the solution is numerically consistent across multiple trials.

\section{Simulation Studies}\label{simulation}

\begin{table}[htpb]
\centering
\caption{Parameter Values Across Simulation Scenarios}
\label{simparams}
\begin{tabular}{lcccc}
\hline
Parameter & Scenario 1 & Scenario 2 & Scenario 3 & Scenario 4 \\
\hline
Class Distinction ($\eta$) & Strong & Weak & Strong & Weak \\
Profile Balance ($\gamma$) & Equal & Equal & Unequal & Unequal \\
\hline
$\eta_{1_1|1}$ & 0.7 & 0.2 & 0.7 & 0.2 \\
$\eta_{2_1|1}$ & 0.1 & 0.1 & 0.1 & 0.1 \\
$\eta_{3_1|1}$ & 0.1 & 0.2 & 0.1 & 0.2 \\
$\eta_{4_1|1}$ & 0.1 & 0.5 & 0.1 & 0.5 \\
$\eta_{1_2|1}$ & 0.1 & 0.5 & 0.1 & 0.5 \\
$\eta_{2_2|1}$ & 0.1 & 0.2 & 0.1 & 0.2 \\
$\eta_{3_2|1}$ & 0.7 & 0.1 & 0.7 & 0.1 \\
$\eta_{4_2|1}$ & 0.1 & 0.2 & 0.1 & 0.2 \\
$\eta_{1_3|1}$ & 0.1 & 0.2 & 0.1 & 0.2 \\
$\eta_{2_3|1}$ & 0.1 & 0.5 & 0.1 & 0.5 \\
$\eta_{3_3|1}$ & 0.1 & 0.2 & 0.1 & 0.2 \\
$\eta_{4_3|1}$ & 0.7 & 0.1 & 0.7 & 0.1 \\
\hline
$\eta_{1_1|2}$ & 0.1 & 0.1 & 0.1 & 0.1 \\
$\eta_{2_1|2}$ & 0.1 & 0.2 & 0.1 & 0.2 \\
$\eta_{3_1|2}$ & 0.7 & 0.5 & 0.7 & 0.5 \\
$\eta_{4_1|2}$ & 0.1 & 0.2 & 0.1 & 0.2 \\
$\eta_{1_2|2}$ & 0.1 & 0.2 & 0.1 & 0.2 \\
$\eta_{2_2|2}$ & 0.7 & 0.1 & 0.7 & 0.1 \\
$\eta_{3_2|2}$ & 0.1 & 0.2 & 0.1 & 0.2 \\
$\eta_{4_2|2}$ & 0.1 & 0.5 & 0.1 & 0.5 \\
$\eta_{1_3|2}$ & 0.7 & 0.5 & 0.7 & 0.5 \\
$\eta_{2_3|2}$ & 0.1 & 0.2 & 0.1 & 0.2 \\
$\eta_{3_3|2}$ & 0.1 & 0.1 & 0.1 & 0.1 \\
$\eta_{4_3|2}$ & 0.1 & 0.2 & 0.1 & 0.1 \\
\hline
$\gamma_{1}$ & 0.5 & 0.5 & 0.3 & 0.3 \\
$\gamma_{2}$ & 0.5 & 0.5 & 0.7 & 0.7 \\
\hline
\end{tabular}
\end{table}

\subsection{Simulation Design}

\noindent To evaluate the performance of the proposed Mixed-LCPM, we conducted extensive simulation studies under varying scenarios. We generated 900 datasets per scenario, varying sample sizes ($N \in \{250, 500, 1000\}$). Each dataset consisted of $N$ subjects observed over $T=3$ time points, with $K=4$ latent classes and $S=2$ latent profiles. Each synthetic data set was simulated to mimic the real data example that we employed in Section \ref{ADNI}. Specifically, Each data set consists of three continuous outcome variables $[Y_{1t},\dots,Y_{3t}]$ and four binary outcome variables $[Z_{1t},\dots,Z_{4t}]$ measured over three time periods: $t = 1, 2, 3$. Continuous outcomes were generated from a multivariate normal distribution with class-specific mean vectors and covariance matrices as follows: $[\textbf{Y}_{t} \mid C_t = c] \sim \textit{MVN}_3(\mu_{c}, \Sigma_c)$, with identity covariance $\Sigma_c = cI_3$, for $c=1,\dots,4$. The class-specific mean vectors were time-invariant and defined as:
\[
\mu_{1} = [4, 0, 0]^\top, \quad \mu_{2} = [-4, 0, 0]^\top, \quad \mu_{3} = [0, 4, 4]^\top, \quad \mu_{4} = [0, 4, -4]^\top.
\]
Similarly, categorical outcomes were generated from a multinomial distribution with class-specific probabilities as follows:
\[ 
[Z_{ipt} \mid C_t = c] \sim \mathrm{Multinomial}(1; \pi_{p1|c}, \pi_{p2|c}), 
\]
where probabilities were set to discriminate between classes (e.g., $\pi_{p|c} \in \{0.9, 0.1\}$), consistent across time points.

\noindent We examined four scenarios (Table \ref{simparams}) designed to test model performance under varying degrees of class separation (Strong vs. Weak) and profile balance (Equal vs. Unequal). All parameters not mentioned in Table \ref{simparams} remained constant throughout. We also ran each scenario with and without the addition of covariates. The scenarios without covariates focused on varying either $\eta$ or $\gamma$ or both, while keeping other parameters constant, in order to investigate the effect of the strength of the separation of classes and/or trajectories. 

\begin{enumerate}
    \item \textbf{Class Separation ($\eta$):} "Strong distinction" implies high probabilities in the class assignment ($\eta \approx 0.7$), while "Weaker distinction" implies higher mixing ($\eta \approx 0.5$).
    \item \textbf{Profile Balance ($\gamma$):} "Equal profiles" assumes a 50/50 split between profiles ($\gamma=[0.5, 0.5]$). "Unequal profiles" assumes a 70/30 split.
\end{enumerate}

\noindent For scenarios including covariates, we simulated a single continuous covariate $X_i \sim \mathcal{N}(1,1)$. To achieve the desired profile proportions, we specified the logistic regression coefficients $\beta$ for Profile 2 (relative to reference Profile 1) as $\beta_2 = [-1, 1]^\top$ for the balanced case (50/50) and $\beta_2 = [-1, 2]^\top$ for the unbalanced case (30/70).

\subsection{Simulation Results}

\begin{figure}[h!]
    \caption{Coverage Probabilities of All Parameters \label{coverage}}
    \centering
    \includegraphics[width=0.99\columnwidth]{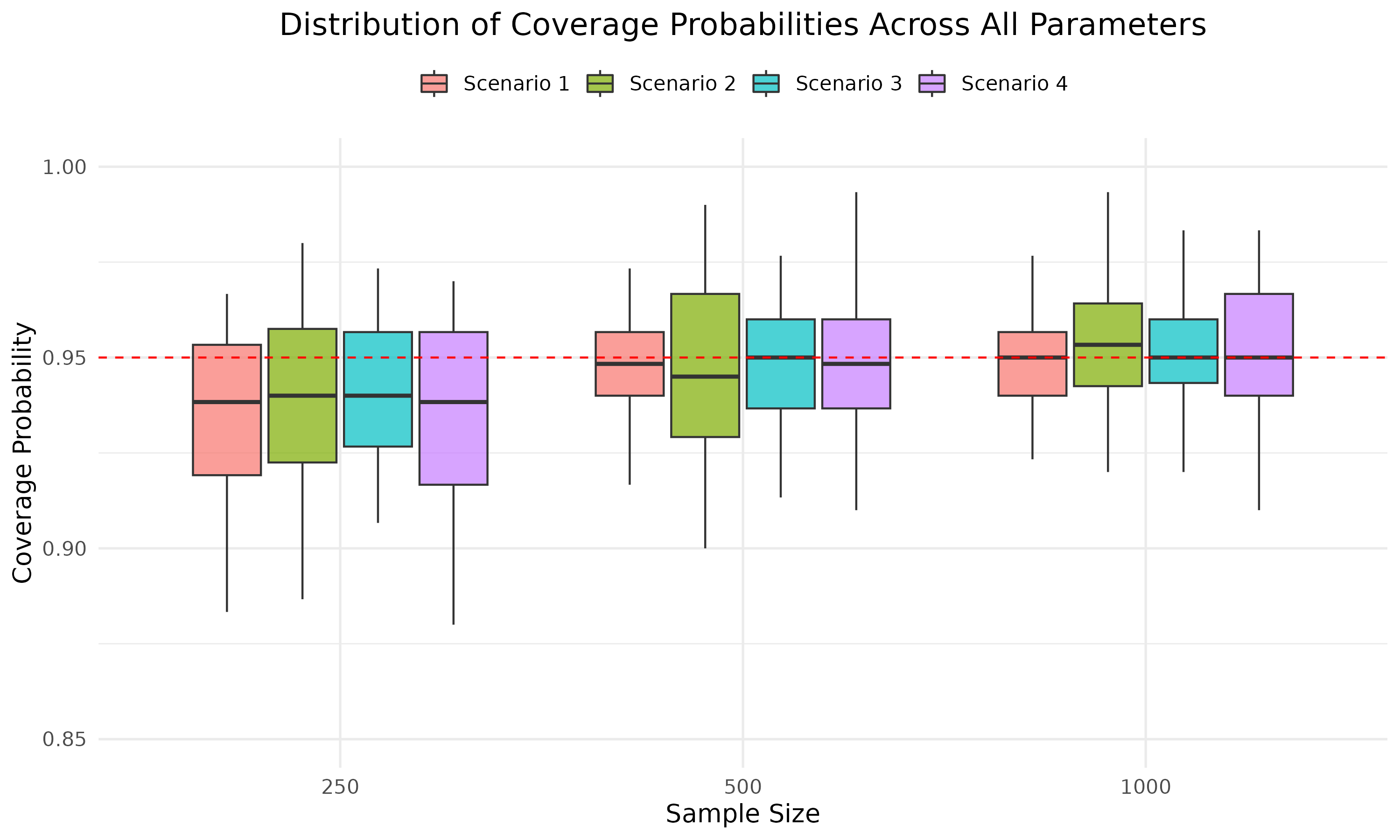}
\end{figure}

In each scenario, we simulated $900$ data sets using the true parameters, fit the Mixed-LCPM, and then calculated the point and interval estimates. We evaluated model performance using standardized bias (Sbias), root mean-squared error (RMSE), and coverage probability (CP) of the 95\% confidence intervals. Table \ref{allscenarios} presents the results for $N=1,000$ across all four scenarios. Comprehensive results for all sample sizes and parameter configurations are provided in the Supplementary Material. In general, the estimator demonstrated robust performance. For $N=1,000$, coverage probabilities consistently remained near the nominal 0.95 level, and standardized bias remained minimal ($|\text{Sbias}| < 0.1$).  We observed that coverage probabilities were sensitive to sample size. At $N=250$, coverage for certain parameters dropped below 0.90 (see Supplementary Material). However, this is expected behavior in mixture modeling due to the asymptotic nature of the standard errors. As sample size increased to $N=500$ and $N=1,000$, coverage improved and RMSE decreased, confirming the consistency of the estimator. Further, for brevity, we only present results on parameters of interest, but we have evaluated performance on all parameters. Figure \ref{coverage} illustrates the distribution of all coverage probabilities, highlighting that the median coverage centers on $0.95$ across all scenarios. Based on these results, we conclude that the proposed estimation algorithm is valid and feasible for the sample sizes typically observed in longitudinal clinical studies.

\begin{table}[ht!]
\centering
\caption{Coverage, RMSE, and Sbias for all scenarios at $N=1,000$}
\label{allscenarios}
\begin{tabular}{l|ccc|ccc}
\hline
Param & \multicolumn{3}{c}{Scenario 1} & \multicolumn{3}{c}{Scenario 2} \\
 & Coverage & RMSE & Sbias & Coverage & RMSE & Sbias \\
\hline
$\gamma_1(\textbf{x}_i)$ & 0.947 & 2.004 &  0.041 & 0.930 & 2.046 & -0.039 \\
$\eta_{1_1|1}$ & 0.943 & 1.551 & -0.017 & 0.953 & 1.543 &  0.035 \\
$\eta_{1_1|2}$ & 0.940 & 1.511 & -0.085 & 0.937 & 1.483 &  0.057 \\
$\eta_{1_2|1}$ & 0.950 & 1.443 & -0.127 & 0.950 & 1.416 &  0.001 \\
$\eta_{1_2|2}$ & 0.923 & 1.468 & -0.005 & 0.907 & 1.434 &  0.011 \\
$\eta_{1_3|1}$ & 0.937 & 1.425 &  0.018 & 0.937 & 1.436 & -0.142 \\
$\eta_{1_3|2}$ & 0.940 & 1.440 & -0.054 & 0.947 & 1.416 &  0.029 \\
$\eta_{2_1|1}$ & 0.953 & 1.492 &  0.107 & 0.957 & 1.483 &  0.062 \\
$\eta_{2_1|2}$ & 0.927 & 1.444 & -0.028 & 0.930 & 1.435 & -0.027 \\
$\eta_{2_2|1}$ & 0.953 & 1.475 & -0.024 & 0.943 & 1.440 &  0.066 \\
$\eta_{2_2|2}$ & 0.937 & 1.510 & -0.025 & 0.920 & 1.487 & -0.065 \\
$\eta_{2_3|1}$ & 0.947 & 1.442 & -0.010 & 0.953 & 1.414 & -0.101 \\
$\eta_{2_3|2}$ & 0.933 & 1.465 & -0.076 & 0.953 & 1.438 &  0.116 \\
$\eta_{3_1|1}$ & 0.953 & 1.443 & -0.030 & 0.933 & 1.438 & -0.012 \\
$\eta_{3_1|2}$ & 0.973 & 1.442 &  0.101 & 0.957 & 1.415 &  0.041 \\
$\eta_{3_2|1}$ & 0.950 & 1.511 & -0.047 & 0.953 & 1.485 &  0.131 \\
$\eta_{3_2|2}$ & 0.940 & 1.432 &  0.141 & 0.937 & 1.437 & -0.113 \\
$\eta_{3_3|1}$ & 0.950 & 1.476 &  0.091 & 0.940 & 1.441 &  0.042 \\
$\eta_{3_3|2}$ & 0.950 & 1.510 &  0.147 & 0.927 & 1.484 & -0.073 \\
\hline
\end{tabular}

\vspace{0.5cm}
\begin{tabular}{l|ccc|ccc}
\hline
Param & \multicolumn{3}{c}{Scenario 3} & \multicolumn{3}{c}{Scenario 4} \\
 & Coverage & RMSE & Sbias & Coverage & RMSE & Sbias \\
\hline
$\gamma_1(\textbf{x}_i)$ & 0.960 & 2.018 &  0.048 & 0.910 & 2.043 & -0.030 \\
$\eta_{1_1|1}$ & 0.963 & 2.139 &  0.027 & 0.940 & 2.227 &  0.144 \\
$\eta_{1_1|2}$ & 0.947 & 1.531 &  0.015 & 0.947 & 1.502 &  0.147 \\
$\eta_{1_2|1}$ & 0.927 & 1.467 & -0.149 & 0.957 & 1.438 & -0.139 \\
$\eta_{1_2|2}$ & 0.947 & 1.477 & -0.051 & 0.933 & 1.453 & -0.018 \\
$\eta_{1_3|1}$ & 0.943 & 1.438 &  0.032 & 0.933 & 1.457 & -0.099 \\
$\eta_{1_3|2}$ & 0.950 & 1.464 &  0.040 & 0.947 & 1.435 & -0.022 \\
$\eta_{2_1|1}$ & 0.937 & 1.502 &  0.098 & 0.967 & 1.500 &  0.120 \\
$\eta_{2_1|2}$ & 0.943 & 1.462 &  0.036 & 0.950 & 1.455 & -0.097 \\
$\eta_{2_2|1}$ & 0.947 & 1.495 & -0.010 & 0.943 & 1.449 &  0.076 \\
$\eta_{2_2|2}$ & 0.970 & 1.529 & -0.089 & 0.940 & 1.502 &  0.074 \\
$\eta_{2_3|1}$ & 0.947 & 1.463 & -0.064 & 0.953 & 1.438 & -0.144 \\
$\eta_{2_3|2}$ & 0.943 & 1.475 & -0.105 & 0.947 & 1.452 &  0.026 \\
$\eta_{3_1|1}$ & 0.940 & 1.461 &  0.006 & 0.933 & 1.454 & -0.026 \\
$\eta_{3_1|2}$ & 0.933 & 1.463 &  0.116 & 0.933 & 1.436 &  0.098 \\
$\eta_{3_2|1}$ & 0.937 & 1.531 & -0.055 & 0.947 & 1.502 &  0.004 \\
$\eta_{3_2|2}$ & 0.953 & 1.438 &  0.100 & 0.927 & 1.455 & -0.081 \\
$\eta_{3_3|1}$ & 0.947 & 1.495 &  0.048 & 0.943 & 1.451 &  0.052 \\
$\eta_{3_3|2}$ & 0.960 & 1.528 &  0.083 & 0.930 & 1.503 &  0.009 \\
\hline
\end{tabular}
\end{table}

\section{Application}\label{ADNI}

\subsection{Data Description}

\noindent We applied the proposed Mixed-LCPM to data obtained from the Alzheimer’s Disease Neuroimaging Initiative (ADNI) database (adni.loni.usc.edu). Launched in 2003, ADNI is a public-private partnership led by Principal Investigator Michael W. Weiner, MD aimed at testing whether serial MRI, PET, biological markers, and clinical assessments can be combined to measure the progression of mild cognitive impairment (MCI) and early Alzheimer’s disease (AD). 

\noindent The continuous outcomes chosen for modeling include composite cognitive measures for memory, language, and executive function harmonized by the Alzheimer's Disease Sequencing Project Phenotype Harmonization Consortium (ADSP‐PHC) (\citep{national_alzheimers_project_act_alzheimers_nodate}).
In addition to these continuous measures, we included three derived categorical variables: Blood Pressure (BP) Level, dichotomized as "High" if systolic/diastolic measurements exceeded 120/80 mmHg at a visit; BMI, dichotomized as "Overweight/Obese" if BMI $\geq 25$; and Clinical Dementia Rating (CDR), dichotomized as "Impaired" (CDR $> 0$) or "Cognitively Normal" (CDR $= 0$). We analyzed six time points: Baseline (Month 0), and Months 6, 12, 24, 36, and 48. Our primary research goal was to identify distinct subgroups of individuals sharing similar joint trajectories of cognitive functioning and vascular risk factors over time.

\noindent Model selection in latent variable modeling is an iterative process requiring the simultaneous specification of the number of latent classes ($K$) and latent profiles ($S$). We performed a grid search ranging $K \in [2, 6]$ and $S \in [2, 6]$, resulting in 25 candidate models. To ensure global maxima were reached, each model was fit using 100 random initial starts with a convergence criterion of a change in log-likelihood $< 10^{-5}$ (max iterations = 500).

\noindent Upon identifying the optimal $[K, S]$ structure based on BIC, we refitted the final model including covariates. Further, for estimates located close to the boundary, we set those parameters to a fixed value of 0 or 1 to reduce the total number of parameters. To ensure interpretability, latent classes and profiles were reordered such that Class 1 and Profile 1 represented the "healthiest" states (highest cognitive scores, lowest risk probabilities). We calculated Odds Ratios (OR) and $95\%$ Confidence Intervals (CI) to quantify the association between baseline covariates and trajectory membership, utilizing Profile 1 as the reference group. Standard errors were obtained by inverting the Hessian matrix.

\begin{table}[ht]
\centering
\caption{Baseline statistics for the overall sample.}
\begin{tabular}{rl}
\hline
 & Baseline Value \\ 
\hline
 & N = 919 \\ 
Age (mean [SD]) & 73.42 (6.98) \\ 
Gender Male (n [\%]) & 515 (56.0) \\  
High Blood Pressure (n [\%]) & 731 (79.8) \\ 
BMI Overweight/Obese (n [\%]) & 605 (66.0) \\
CDR Impaired  (n [\%])  & 599 (65.2) \\ 
Number of APOE4 Alleles (n [\%])  & \\ 
\hspace{1em} 0 & 539 (58.7) \\ 
\hspace{1em} 1 & 305 (33.2) \\ 
\hspace{1em} 2 & 75 (8.2) \\ 
Years of Education (mean [SD]) & 16.11 (2.74) \\ 
Memory Score (mean [SD]) & 0.37 (0.61) \\ 
Executive Function Score (mean [SD]) & 0.55 (0.60) \\ 
Language Score (mean [SD]) & 0.54 (0.55) \\ 
Visuospatial Score (mean [SD]) & 0.40 (0.54) \\ 
Number of Visits (mean [SD]) & 5.54 (0.50) \\
\hline
\end{tabular}
\label{tab:baseline}
\end{table}

\subsection{Results}
The final analytic sample included $N=919$ participants (Table \ref{tab:baseline}). The optimal model structure was identified as 5 Latent Classes and 6 Latent Profiles. Table \ref{tab:classlevelresults} details the characteristics of the five latent classes. Positive mean values for cognitive domains indicate better-than-average functioning. Classes 1 and 2 both exhibit high cognitive scores, indicating normal cognition. However, they diverge significantly regarding vascular risk; Class 1 is characterized by low probabilities of both hypertension and high BMI (labeled ``Normal \& Healthy''), whereas Class 2 exhibits high probabilities of these risk factors (``Normal \& Unhealthy''). Similarly, Classes 3 and 4 both show reduced cognitive scores consistent with mild impairment (CDR $>0$ probabilities $>0.95$). Class 3 carries a high vascular burden (``Mild Impairment \& Unhealthy''), while Class 4 presents with a healthier vascular profile (``Mild Impairment \& Healthy''). Finally, Class 5 exhibits the lowest cognitive scores across all domains combined with high rates of vascular risk factors, representing a ``Significant Impairment'' state.

\begin{table}[!htb]
\centering
\begin{minipage}{\columnwidth}
\centering
\caption{Means and Probabilities by Latent Class}
\label{tab:classlevelresults}
\begin{tabular}{lrrrrr}
\toprule
\textbf{Variable} & \textbf{Class 1} & \textbf{Class 2} & \textbf{Class 3} & \textbf{Class 4} & \textbf{Class 5} \\
\midrule
\multicolumn{6}{l}{\textbf{Cognitive Variables (Means)}} \\
Memory & 0.911 & 0.951 & -0.050 & -0.170 & -0.941 \\
Executive Function & 0.880 & 0.888 & 0.329 & 0.445 & -0.859 \\
Language & 0.847 & 0.896 & 0.308 & 0.288 & -0.624 \\
Visuospatial & 0.535 & 0.539 & 0.289 & 0.427 & -0.581 \\
\midrule
\multicolumn{6}{l}{\textbf{Categorical Variables (Probabilities)}} \\
High BP & 0.657 & 0.808 & 0.822 & 0.720 & 0.738 \\
High BMI & 0.026 & $1.00^\dagger$ & $1.00^\dagger$ & 0.095 & 0.514 \\
CDR $>$ 0 & 0.258 & 0.305 & 0.951 & 0.977 & $1.00^\dagger$ \\
\bottomrule
\end{tabular}%
\vspace{1ex}

{\footnotesize
\raggedright
\textit{Notes:}
$\dagger$ Indicates parameters constrained to 1 or 0. 
Abbreviations: BP = blood pressure; BMI = body mass index; CDR = Clinical Dementia Rating.
}
\end{minipage}
\end{table}

\begin{figure}[htbp]
    \caption{\label{fig:trajectories} Marginal Trends of Outcome Variables Over Time}
    \centering
    \includegraphics[width=\columnwidth]{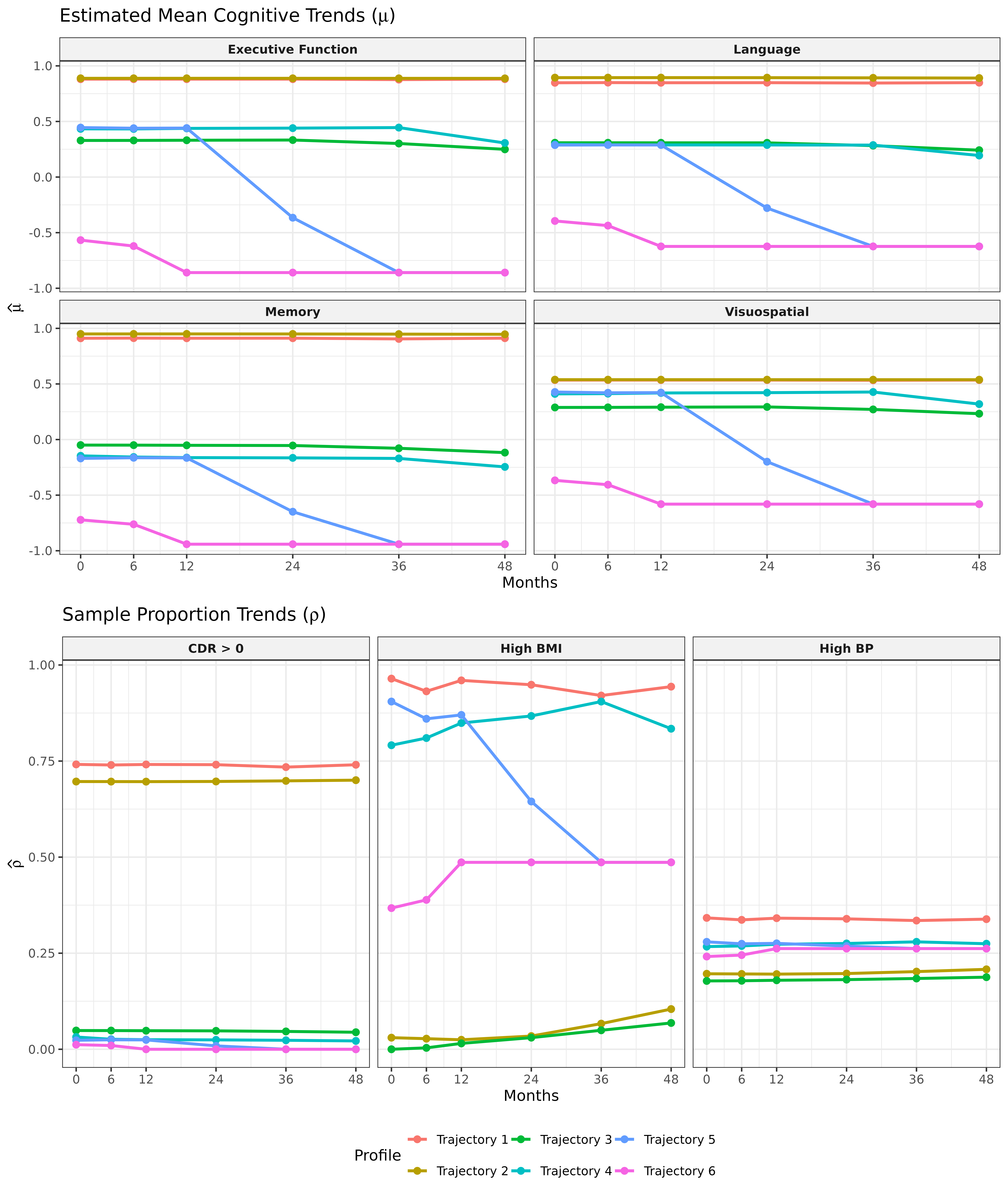}
\end{figure}

\noindent After examining the class-level values, we examine the secondary measurement parameters, which represent the probability of being in a specific class at a specific time point, as shown in Table \ref{tab:tab:trajcetoryprobs}. The marginal mean trends are visualized in Figure \ref{fig:trajectories}. The first latent trajectory, which accounts for about $16\%$ of the sample, represents patients who remain in the Cognitively Normal and Healthy class across all time points. A similar pattern in which the class at time 1 remains steady across all $6$ time periods is exhibited across all trajectories except Trajectory 5. Here, Trajectory 5 represents a group of participants who are originate in class 4 (Mild Impairment and Healthy) but transition to the Significant Impairment class with high probability by time 5. We name the Profiles (1) Steady Cognition and Steady Good Health; (2) Steady Cognition and Steady Bad Health; (3) Steady Mild Impairment and Steady Bad Health; (4) Steady Mild Impairment and Steady Good Health; (5) Progressive Cognitive Decline and Worsening Health; (6) Steady Cognitive Impairment and Unhealthy. Note that the naming conventions are subjective and highly dependent on the clinical/domain knowledge. 

\noindent The marginal trends of outcomes presented in Figure \ref{fig:trajectories} clearly indicate the novelty of the proposed Mixed-LCPM; it allows us to identify a larger number of latent trajectories that are distinguished by their behaviors that would have been disguised if a univariate or a single-type trajectory analysis were implemented. For example, if a trajectory analysis only employed four cognitive scores, the first two profiles would have been considered as a single trend, even though subjects in Profiles (1) and (2) are noticeably different in their blood pressure trends. Similarly, although subjects in Profiles (3) and (4) present almost similar longitudinal behaviors on four cognitive scores, their longitudinal BMI are noticeably distinguished; while subjects in Profile (3) are highly overweight, subjects in Profile (4) are unlikely to be overweight. Therefore, by modeling the joint trajectories of both the continuous and categorical outcomes, we can identify distinct multidimensional profiles that capture the complex heterogeneity of disease progression, revealing subgroup differences that would otherwise remain obscured in univariate analyses.

\begin{table}[htbp]
\centering
\begin{minipage}{\columnwidth}
\centering
\caption{\label{tab:tab:trajcetoryprobs}Class Probabilities by Trajectory}
\centering
\resizebox{\ifdim\width>\linewidth\linewidth\else\width\fi}{!}{
\begin{tabular}[t]{lrrrrrrr}
\toprule
Trajectory & Class & Time 1 & Time 2 & Time 3 & Time 4 & Time 5 & Time 6\\
\midrule
 & Healthy & \textbf{0.990} & \textbf{0.956} & \textbf{0.985} & \textbf{0.974} & \textbf{0.945} & \textbf{0.969}\\
 & Cog Normal/ Unhealthy & 0.010 & 0.044 & 0.015 & 0.026 & 0.048 & 0.031\\
Trajectory 1 & Questionable Cog \& Unhealthy & $0.00^\dagger$ & $0.00^\dagger$ & $0.00^\dagger$ & $0.00^\dagger$ & 0.008 & $0.00^\dagger$\\
 (15.9\%) & Questionable Cog \& Healthy & $0.00^\dagger$ & $0.00^\dagger$ & $0.00^\dagger$ & $0.00^\dagger$ & $0.00^\dagger$ & $0.00^\dagger$\\
 & Cognitively Impaired & $0.00^\dagger$ & $0.00^\dagger$ & $0.00^\dagger$ & $0.00^\dagger$ & $0.00^\dagger$ & $0.00^\dagger$\\
\addlinespace
& Healthy  & 0.031 & 0.028 & 0.025 & 0.035 & 0.068 & 0.107\\
& Cog Normal/ Unhealthy & \textbf{0.969} & \textbf{0.972} & \textbf{0.975} & \textbf{0.965} & \textbf{0.932} & \textbf{0.893}\\
Trajectory 2 &  Questionable Cog \& Unhealthy & $0.00^\dagger$ & $0.00^\dagger$ & $0.00^\dagger$ & $0.00^\dagger$ & $0.00^\dagger$ & $0.00^\dagger$\\
(32.3\%) & Questionable Cog \& Healthy & $0.00^\dagger$ & $0.00^\dagger$ & $0.00^\dagger$ & $0.00^\dagger$ & $0.00^\dagger$ & $0.00^\dagger$\\
& Cognitively Impaired & $0.00^\dagger$ & $0.00^\dagger$ & $0.00^\dagger$ & $0.00^\dagger$ & $0.00^\dagger$ & $0.00^\dagger$\\
\addlinespace
& Healthy & $0.00^\dagger$ & $0.00^\dagger$ & $0.00^\dagger$ & $0.00^\dagger$ & $0.00^\dagger$ & $0.00^\dagger$\\
& Cog Normal/ Unhealthy & $0.00^\dagger$ & $0.00^\dagger$ & $0.00^\dagger$ & $0.00^\dagger$ & $0.00^\dagger$ & $0.00^\dagger$\\
Trajectory 3 & Questionable Cog \& Unhealthy & \textbf{1.000} & \textbf{0.996} & \textbf{0.983} & \textbf{0.967} & \textbf{0.933} & \textbf{0.892}\\
(26.9\%) & Questionable Cog \& Healthy & $0.00^\dagger$ & 0.004 & 0.017 & 0.033 & 0.040 & 0.038\\
& Cognitively Impaired & $0.00^\dagger$ & $0.00^\dagger$ & $0.00^\dagger$ & $0.00^\dagger$ & 0.027 & 0.071\\
\addlinespace
& Healthy & 0.007 & $0.00^\dagger$ & $0.00^\dagger$ & $0.00^\dagger$ & $0.00^\dagger$ & $0.00^\dagger$\\
& Cog Normal/ Unhealthy & $0.00^\dagger$ & $0.00^\dagger$ & $0.00^\dagger$ & $0.00^\dagger$ & $0.00^\dagger$ & $0.00^\dagger$\\
Trajectory 4 & Questionable Cog \& Unhealthy & 0.126 & 0.105 & 0.062 & 0.041 & $0.00^\dagger$ & 0.030\\
(14.3\%) & Questionable Cog \& Healthy & \textbf{0.866} & \textbf{0.895} & \textbf{0.938} & \textbf{0.959} & \textbf{1.00}$^\dagger$ & \textbf{0.866}\\
& Cognitively Impaired & $0.00^\dagger$ & $0.00^\dagger$ & $0.00^\dagger$ & $0.00^\dagger$ & $0.00^\dagger$ & 0.104\\
\addlinespace
 & Healthy & $0.00^\dagger$ & $0.00^\dagger$ & $0.00^\dagger$ & $0.00^\dagger$ & $0.00^\dagger$ & $0.00^\dagger$\\
 & Cog Normal/ Unhealthy & $0.00^\dagger$ & $0.00^\dagger$ & $0.00^\dagger$ & $0.00^\dagger$ & $0.00^\dagger$ & $0.00^\dagger$\\
Trajectory 5 & Questionable Cog \& Unhealthy & $0.00^\dagger$ & 0.050 & 0.038 & $0.00^\dagger$ & $0.00^\dagger$ & $0.00^\dagger$\\
(4.5\%) & Questionable Cog \& Healthy & \textbf{1.00}$^\dagger$ & \textbf{0.950} & \textbf{0.962} & 0.378 & $0.00^\dagger$ & $0.00^\dagger$\\
& Cognitively Impaired & $0.00^\dagger$ & $0.00^\dagger$ & $0.00^\dagger$ & \textbf{0.622} & \textbf{1.00}$^\dagger$ & \textbf{1.00}$^\dagger$\\
\addlinespace
& Healthy & $0.00^\dagger$ & $0.00^\dagger$ & $0.00^\dagger$ & $0.00^\dagger$ & $0.00^\dagger$ & $0.00^\dagger$\\
 & Cog Normal/ Unhealthy & $0.00^\dagger$ & $0.00^\dagger$ & $0.00^\dagger$ & $0.00^\dagger$ & $0.00^\dagger$ & $0.00^\dagger$\\
Trajectory 6 & Questionable Cog \& Unhealthy & 0.245 & 0.201 & $0.00^\dagger$ & $0.00^\dagger$ & $0.00^\dagger$ & $0.00^\dagger$\\
(6.2\%) & Questionable Cog \& Healthy & $0.00^\dagger$ & $0.00^\dagger$ & $0.00^\dagger$ & $0.00^\dagger$ & $0.00^\dagger$ & $0.00^\dagger$\\
& Cognitively Impaired & \textbf{0.754} & \textbf{0.799} & \textbf{1.00}$^\dagger$ & \textbf{1.00}$^\dagger$ & \textbf{1.00}$^\dagger$ & \textbf{1.00}$^\dagger$\\
\bottomrule
\end{tabular}}
\vspace{1ex}

{\footnotesize
\raggedright
\textit{Note:}
$\dagger$ Indicates parameters constrained to 1 or 0.
}
\end{minipage}
\end{table}

\begin{figure}[htbp]
    \caption{\label{fig:orci}}
    \centering
    \includegraphics[width=0.99\columnwidth]{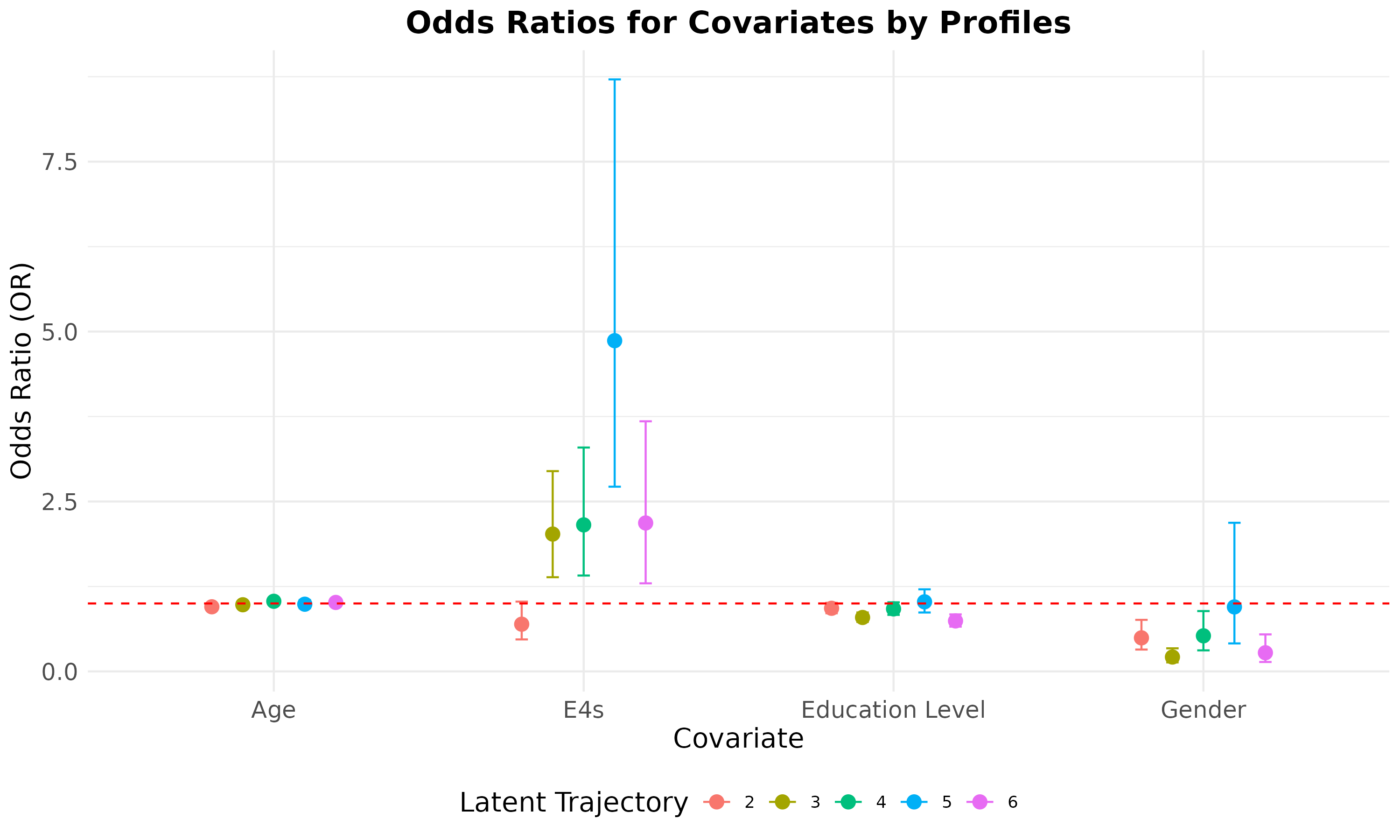}
\end{figure}

\noindent Finally, Table \ref{tab:or_ci} (visualized in Figure \ref{fig:orci}) displays the association between baseline covariates and profile membership, relative to the reference Trajectory 1 (Normal and Healthy). For example, regarding Age, the odds of belonging to Profile 2 (Normal and Unhealthy) compared to Profile 1 decrease by approximately 5\% for every one-year increase in age ($\text{OR} = 0.95$, 95\% CI $[0.92, 0.98]$).

\begin{table}[htbp]
\centering
\caption{\label{tab:or_ci}Odds Ratios and 95\% Confidence Intervals by Covariate for Each Latent Trajectory (Trajectory 1 as Baseline)}
\begin{tabular}{lccccc}
\toprule
Profile & Age & Gender (F v M) & Education Level & Number of E4s \\
\midrule
Profile 2 & 0.951 (0.923, 0.980) & 0.494 (0.322, 0.759) & 0.929 (0.856, 1.009) & 0.695 (0.471, 1.027) \\
Profile 3 & 0.981 (0.949, 1.013) & 0.212 (0.133, 0.340) & 0.794 (0.728, 0.867) & 2.021 (1.386, 2.947) \\
Profile 4 & 1.032 (0.994, 1.072) & 0.524 (0.310, 0.888) & 0.920 (0.832, 1.017) & 2.156 (1.411, 3.294) \\
Profile 5 & 0.989 (0.931, 1.050) & 0.949 (0.412, 2.187) & 1.024 (0.867, 1.208) & 4.866 (2.718, 8.710) \\
Profile 6 & 1.015 (0.965, 1.067) & 0.275 (0.139, 0.545) & 0.743 (0.659, 0.839) & 2.184 (1.296, 3.680) \\
\bottomrule
\end{tabular}
\end{table}

\section{Discussion}

This paper proposes an advanced latent class trajectory model that identifies latent trajectories and summarizes the longitudinal behaviors of subjects with multivariate mixed outcomes. The proposed model is a natural extension of the LCPM \citep {chung_latent_2011, lee_latent_2025}, which has been developed to identify latent trajectories of populations based on repeatedly measured multiple categorical outcomes. This paper expands the idea of LCPM by allowing the vector of multivariate outcomes to be mixed with both categorical and continuous variables. Specifically, the proposed LCPM introduces two types of categorical latent variables: a vector of categorical latent variables, each summarizing the response patterns of the outcome variables at a time point, and a categorical latent trajectory variable that summarizes the sequential patterns of latent classes. In this manner, subjects with the same latent trajectory membership present common latent class patterns across time periods and thus share similar longitudinal outcome patterns. In addition, associations between individual-level covariates and the latent trajectory memberships are estimated via a multinomial logistic regression model embedded within the trajectory model, providing interpretable insights into how covariates relate to longitudinal outcome trends via odds-ratio scales.

\noindent Unlike traditional trajectory models that describe longitudinal trajectories of outcome via parametric functions of time and covariates, our proposed model identifies latent trajectories of outcomes in a non-parametric manner; the latent class variable at each time point describes representative outcome patterns, and the latent trajectory variable describes representative patterns of how individual-level latent class memberships transit over time. Such a two-level latent structure flexibly describes the longitudinal behaviors of multivariate outcomes without specifying the functional form of latent trajectories.  

\noindent In this paper, we employed the EM algorithm to obtain ML estimates of the model parameters and discuss the implementation details in the Supplemental Material. Associations between latent trajectory memberships and subject-level covariates were estimated via one-step estimation, in which the measurement parameters and regression coefficients are simultaneously estimated. The standard errors of the ML estimates are obtained from the inverse of the Hessian matrix evaluated at the ML estimates. Through the extensive simulation studies, we confirmed that the proposed EM algorithm stably yields unbiased estimates and proper $95\%$ confidence intervals for all measurement and structural parameters. To promote our methodology, we have written a program to implement LCPM in the R language (version 4.4.1) available at {\em https://github.com/lindsaysalvati/thesis} or upon request. 

\noindent We acknowledge two drawbacks of the proposed estimation strategy. First, the proposed EM algorithm fails to converge to the global maximum if it starts at an inappropriate initial value. To avoid the local-maxima problem, we used $100$ initial values and selected the result with the highest log-likelihood as the final solution. As an alternative approach, Bayesian estimation using random draws from the posterior distributions of the parameters, given the data and the prior distribution, is under investigation. Second, the number of latent classes and trajectories must be determined by the analyst. This implies that fitting multiple candidate models with varying numbers of latent classes and profiles is required, which entails heavy computation, as each candidate model must be fit with different initial values. Unfortunately, such computational costs cannot be avoided when using maximum likelihood estimation. As an alternative approach, introducing a prior distribution over the numbers of latent classes and trajectories using the Dirichlet Process is under investigation. 


\noindent To demonstrate the use of the proposed trajectory analysis, we employed data from the Alzheimer’s Disease Neuroimaging Initiative (ADNI) to investigate whether distinct subgroups of individuals shared similar joint trajectories of cognitive functioning and health-related risk factors over time. Throughout the analysis, we discovered five latent classes that are distinct in behavioral patterns toward four continuous and three categorical outcomes: Healthy, Cognitively Normal/Unhealthy, Mild Impairment \& Healthy, Mild Impairment \& Unhealthy, and Cognitively Impaired. In addition, we identified six latent trajectory groups that capture representative sequential patterns across five latent classes over six time periods. These six trajectories successfully summarize the multivariate, longitudinal behaviors of mixed outcomes in an efficient and interpretable manner; subjects with the same trajectory memberships are homogeneous in their longitudinal patterns toward the latent class memberships, and thus present similar response patterns to the multivariate outcomes at each time point. Lastly, we have discovered significant associations between the latent trajectory memberships and subject-level covariates: Age, Gender, Education level, and the number of APOE4 alleles. In such a way, our proposed trajectory model is capable of discovering (1) representative patterns of multivariate outcomes at each time point, (2) representative longitudinal trajectories of response patterns, and (3) associations between subject-level covariates and the trajectories of the outcomes. Note that these results are based on a particular cohort extracted from the ADNI and thus cannot be used to inform clinical decision-making; they were presented to illustrate the statistical approaches introduced in this article. Such an analysis, performed on carefully recruited data, could inform the risk-benefit analysis for clinical decision-making for patients who are potentially at risk of cognitive impairment or AD. 

\noindent There are two future research directions. First, the proposed model can be extended to clustered data, where subjects are nested within hierarchical units. For example, patient-level outcomes can be collected from multiple independent locations, where patients from the same location are correlated. In such a case, there will be three sources of correlation between measurements: correlations within the same patient at a particular time point, correlations from repeated measurements within the same patient, and correlations between patients from the same location. Such a multi-level structure can be incorporated into a latent trajectory model via an additional latent variable \citep{lee_bayesian_2025}. Second, our proposed model addresses missing values in outcomes under the missing-at-random (MAR) assumption, where missing values are assumed to be related with the observed data only \citep{dempster_maximum_1977}. Specifically, when implementing the EM algorithm, missing values in outcomes are replaced by the conditional expectations of missing values given the observed data and the latent variable memberships \citep{dempster_maximum_1977, lee_latent_2025}. However, occurrences of missing values in longitudinal studies are often related to unobserved values, which implies missing-not-at-random (MNAR). When missing values are MNAR, one should account for the joint distribution of the observed data and the missingness, which can be investigated via joint modeling frameworks, such as pattern mixture models, selection models, or shared parameter models \citep{lee_latent_2024, lee_latent_2025-2}. Developing an extension of the proposed LCPM using the joint modeling framework would be a future research agenda.

\section*{Acknowledgement}

Data collection and sharing for the Alzheimer's Disease Neuroimaging Initiative (ADNI) is funded by the National Institute on Aging (National Institutes of Health Grant U19AG024904). The grantee organization is the Northern California Institute for Research and Education. In the past, ADNI has also received funding from the National Institute of
Biomedical Imaging and Bioengineering, the Canadian Institutes of Health Research, and private sector contributions through the Foundation for the National Institutes of Health (FNIH) including generous contributions from the following: AbbVie, Alzheimer’s Association;
Alzheimer’s Drug Discovery Foundation; Araclon Biotech; BioClinica, Inc.; Biogen; Bristol-Myers Squibb Company; CereSpir, Inc.; Cogstate; Eisai Inc.; Elan Pharmaceuticals, Inc.; Eli Lilly and Company; EuroImmun; F. Hoffmann-La Roche Ltd and its affiliated company
Genentech, Inc.; Fujirebio; GE Healthcare; IXICO Ltd.; Janssen Alzheimer Immunotherapy Research \& Development, LLC.; Johnson \& Johnson Pharmaceutical Research \& Development LLC.; Lumosity; Lundbeck; Merck \& Co., Inc.; Meso Scale Diagnostics, LLC.; NeuroRx Research; Neurotrack Technologies; Novartis Pharmaceuticals Corporation; Pfizer
Inc.; Piramal Imaging; Servier; Takeda Pharmaceutical Company; and Transition Therapeutics.

\section*{Data Availability Statement}

The data that support the findings of this study are available on request from the first author. 

\section*{Conflicts of Interest}

The authors declare no conflicts of interest.

\bibliographystyle{plain}
\bibliography{References.bib}

\appendix

\section{Details of the expectation-maximization (EM) algorithm}

This section presents details of the expectation-maximization (EM) algorithm tailored for the proposed Mixed Latent Class Profile Model (Mixed-LCPM).

\subsection{E-Step}
The focus of the E-step is calculating the expected value of the complete data likelihood $L^\star(u, \textbf{c} | \textbf{y}_i, \textbf{z}_i, \textbf{x}_i)$, which can be written as follows:
\begin{equation}
L^\star(u, \textbf{c} | \textbf{y}_i, \textbf{z}_i, \textbf{x}_i)  =   \gamma_u(\textbf{x}_i) \prod\limits_{t=1}^T \left[ \eta_{c_t|u} ^{(t)} \phi(\textbf{Y}_i| \mu_{c_t,t}, \Sigma_{c_t,t}) 
       \prod\limits_{p=1}^{P_2} \prod\limits_{k=1}^{r_p} \pi_{pk|c_t}^{I(z_{pt}=k)} \right]. 
\end{equation}

\noindent Then, we take the log complete-data likelihood:

\begin{equation}
\begin{aligned}
l^\star(u, \textbf{c} | \textbf{y}, \textbf{z}, \textbf{x}) &= \sum_{i=1}^N \sum_{t=1}^T \sum_{p=1}^{P_2} \sum_{k=1}^{r_p} I(C_t=c_t) I(z_{ipt}=k) \log \pi_{pk|c_t} \\
&\quad + \sum_{i=1}^N \sum_{t=1}^T I(C_t=c_t) \log \phi(\textbf{Y}_i| \mu_{c_t,t}, \Sigma_{c_t,t}) \\
&\quad + \sum_{i=1}^N \sum_{t=1}^T I(U=u,C_t=c_t) \log \eta_{c_t|u}^{(t)} \\
&\quad + \sum_{i=1}^N I(U=u) \log \gamma_u(\textbf{x}_i)
\end{aligned}
\end{equation}
\noindent Lastly, we take the expected value of the latent variables given the observed data $[{\bf Y}]$.

\begin{equation} \label{expected}
    \begin{aligned}
        l^\star(u, \textbf{c} | \textbf{y}, \textbf{z}, \textbf{x}) = \sum\limits_{i=1}^N \sum\limits_{t=1}^T \sum\limits_{p=1}^{P_2} \sum\limits_{k=1}^{r_p} \theta_{i(c_t)} I(z_{ipt}=k) \text{log}\pi_{pk|c_t} \\
        + \sum\limits_{i=1}^N \sum\limits_{t=1}^T \theta_{i(c_t)}  \text{log }\phi(\textbf{Y}_i| \mu_{c_t,t}, \Sigma_{c_t,t}) \\
        +  \sum\limits_{i=1}^N \sum\limits_{t=1}^T \theta_{i(u,c_t)}  \text{log }\eta_{c_t|u} ^{(t)} \\ +  \sum\limits_{i=1}^N \theta_{i(u)} 
        \text{log }\gamma_u(\textbf{x}_i),
    \end{aligned}
\end{equation} 
\noindent where $i=1,\dots,N$, $u=1,\dots,S$, $c_t=1,\dots,K$, and $t=1,\dots,T$ and the marginal probabilities are written as follows:

\begin{equation}\label{fulltheta}
\begin{aligned}
    \theta_{i(u,\textbf{c})}&=P(U=u, \textbf{C}=\textbf{c}|\textbf{y}_i,\textbf{z}_i,\textbf{x}_i) \\
    &= \frac{P(U=u, \textbf{C}=\textbf{c},\textbf{Y}_i=\textbf{y}_i,\textbf{Z}_i=\textbf{z}_i|\textbf{x}_i)}{P(\textbf{Y}_i,\textbf{Z}_i|\textbf{x}_i)} \\
    &= \frac{\gamma_u(\textbf{x}_i) \prod\limits_{t=1}^T \left[ \eta_{c_t|u} ^{(t)} \phi(\textbf{Y}_i| \mu_{c_t,t}, \Sigma_{c_t,t}) 
       \prod\limits_{p=1}^{P_2} \prod\limits_{k=1}^{r_p} \pi_{pk|c_t}^{I(z_{pt}=k)} \right] }{
   \sum\limits_{u=1}^S \sum\limits_{c_1=1}^K \dots \sum\limits_{c_T=1}^K \gamma_u(\textbf{x}_i) \prod\limits_{t=1}^T \left[ \eta_{c_t|u} ^{(t)} \phi(\textbf{Y}_i| \mu_{c_t,t}, \Sigma_{c_t,t}) 
       \prod\limits_{p=1}^{P_2} \prod\limits_{k=1}^{r_p} \pi_{pk|c_t}^{I(z_{pt}=k)} \right] },\\
\theta_{i(u,c_t)}& = 
\sum\limits_{c_1=1}^K \dots \sum\limits_{c_{t-1}=1}^K \sum\limits_{c_{t+1}=1}^K \dots \sum\limits_{c_T=1}^K \theta_{i(u,\textbf{c})} = P( U=u, C_t=c_t|\textbf{y}_i,\textbf{z}_i),\\
\theta_{i(u,c_t,c_{t'})}& = P( U=u, C_t=c_t, C_{t'}=c_{t'}|\textbf{y}_i,\textbf{z}_i,\textbf{x}_i),\\
\theta_{i(u)} &= \sum\limits_{c_t=1}^K \theta_{i(u,c_t)}  = P(U=u|\textbf{y}_i,\textbf{z}_i,\textbf{x}_i),\\
\theta_{i(c_t)} &= \sum\limits_{u=1}^S \theta_{i(u,c_t)} =  P(C_t=c_t) |\textbf{y}_i,\textbf{z}_i,\textbf{x}_i).
\end{aligned}
\end{equation}

\noindent Unfortunately, the computational cost of $\theta_{i(u,\textbf{c})}$ is large, since it is a conditional probability whose dimension is $S\times K^T$, which increases exponentially as $T$ increases. Therefore, to avoid the large computational cost of $\theta_{i(u,\textbf{c})}$, we instead use a forward-backward algorithm. This algorithm decomposes the full conditional probability as follows: \\

\textbf{Forward Probability:} $\alpha_{it}(u,c_t) = P(Y_1=y_1,\ldots,Y_t=y_t, C_t=c_t \mid u) $

=$
\begin{cases}
\eta_{c_t|u}^{(t)} 
\, \phi(\mathbf{Y}_i \mid \mu_{c_t,t}, \Sigma_{c_t,t})
\, \displaystyle\prod_{p=1}^{P_2} 
    \prod_{k=1}^{r_p} 
    \pi_{pk|c_t}^{I(z_{pt}=k)}
\, \displaystyle\sum_{c_{t-1}=1}^{K} 
    \alpha_{i(t-1)}(u,c_{t-1}), 
& \text{if } t \ge 2, \\[10pt]
\eta_{c_1|u}^{(1)} 
\, \phi(\mathbf{Y}_i \mid \mu_{c_1,1}, \Sigma_{c_1,1})
\, \displaystyle\prod_{p=1}^{P_2} 
    \prod_{k=1}^{r_p} 
    \pi_{pk|c_1}^{I(z_{p1}=k)}, 
& \text{if } t = 1.
\end{cases}$

\textbf{Backward Probability:} $\lambda_{it}(u) = P(Y_{t+1}=y_{t+1},\ldots,Y_T=y_T \mid U=u)$

$= \begin{cases}
\displaystyle
\sum_{c_{t+1}=1}^{K} 
  \eta_{c_{t+1}|u}^{(t+1)} 
  \, \phi(\mathbf{Y}_i \mid \mu_{c_{t+1},(t+1)}, \Sigma_{c_{t+1},(t+1)})  
  \prod_{p=1}^{P_2} 
      \prod_{k=1}^{r_p} 
      \pi_{pk|c_{t+1}}^{I(z_{p(t+1)}=k)}
  \, \lambda_{i(t+1)}(u)
, & \text{if } t < T, \\[5pt]
1, & \text{if } t = T.
\end{cases}$\\

\noindent Thus, we can rewrite the conditional probability of belonging to profile $u$ with latent class membership $c_t$ at time $t$ as follows:

\begin{equation}
    \theta_i(u,c_t)= \frac{\gamma_u(\textbf{x}_i)\alpha_{it}(u,c_t)\lambda_{it}(u)}{\sum\limits_{u=1}^S \sum\limits_{c_1=1}^K \dots \sum\limits_{c_T=1}^K \gamma_u(\textbf{x}_i)\alpha_{iT}(u,c_T)}.
\end{equation}

\noindent Finally, let $\hat{\Theta}^{(m)} = [\hat{\pi}^{(m)}, \hat{\eta}^{(m)}, \hat{\mu}^{(m)},\hat{\sigma}^{(m)}]$ be the parameter values at the $m$th iteration. Under the missing at random (MAR) assumption, we can write the conditional expectations in the E-step as: 

\begin{equation}
\begin{aligned}
\hat{Y}_{ipt}^{(m)} =
\begin{cases}
E(\mathbf{Y}_{ipt}^{\text{mis}} \mid \hat{\mathbf{Y}}_{it}^{\text{obs}},\hat{\Theta}^{(m)}), & \text{if } \mathbf{Y}_{ipt} \text{ is missing},\\
\mathbf{Y}_{ipt}, & \text{if } \mathbf{Y}_{ipt} \text{ is observed}.
\end{cases}\\
\hat{I}(\mathbf{Z}_{ipt}=k\mid c_t)^{(m)} =
\begin{cases}
\hat{\pi}^{(m)}_{pk|c}, & \text{if } \mathbf{Z}_{ipt} \text{ is missing},\\
I(\mathbf{Z}_{ipt}=k\mid c_t), & \text{if } \mathbf{Z}_{ipt} \text{ is observed}.
\end{cases}\\
\hat{C}_{i(p,q,c_t,t)}^{(m)} =
\begin{cases}
\displaystyle
\text{Cov}(\mathbf{Y}_{ipt},\mathbf{Y}_{iqt}\mid \hat{\mathbf{Y}}_{it}^{\text{obs}},\hat{\Theta}^{(m)}), & \text{if both } \mathbf{Y}_{ipt} \text{ and } \mathbf{Y}_{iqt} \text{ are missing},\\[2pt]
0, & \text{if either } \mathbf{Y}_{ipt} \text{ or } \mathbf{Y}_{iqt} \text{ is observed}.
\end{cases}
\end{aligned}
\end{equation}

\noindent Intuitively, under MAR assumptions, missing values are replaced by their conditional expected values given latent class memberships and observed data during the E-step.

\subsection{M-Step}

In the M-step, we update the current parameter values using the maximizers of the expected log complete data likelihood from the E-step, as shown in Eq. (\ref{expected}). We use Lagrange multipliers since the sum of parameters that are used in measuring each latent variable is constrained to be 1 as follows:
\begin{align}
\sum\limits_{k=1}^{r_p}\pi_{pk|c_t} = \sum\limits_{c_t=1}^K \eta_{c_t|u} ^{(t)} = \sum\limits_{u=1}^S \gamma_u =1.
\end{align}

\noindent We can update the current parameter values by maximizing Eq.(\ref{expected}) with respect to all parameters, using Lagrange multiplier method as follows:

\begin{equation}\label{Mstep}
\begin{aligned}
\hat{\mu}^{(m+1)}_{p|c}&=\frac{\sum\limits_{i=1}^{N}\sum\limits_{t=1}^{T}\theta_{i(c_t)}\hat{Y}_{ipt}^{(m)}}{\sum\limits_{i=1}^{N}\sum\limits_{t=1}^{T}\theta_{i(c_t)}}, ~~\hat{\pi}^{(m+1)}_{pk|c} = \frac{\sum\limits_{i=1}^{N}\sum\limits_{t=1}^{T}\theta_{i(c_t)}\hat{I}(Z_{ipt}=k|c_t)^{(m)}}{\sum\limits_{i=1}^{N}\sum\limits_{t=1}^{T}\theta_{i(c_t)}}, \\
\hat{\eta}^{(m+1)}_{c_t|u} &= \frac{\sum\limits_{i=1}^{N}\sum\limits_{t=1}^{T}\theta_{i(u,c_t)}}{\sum\limits_{i=1}^{N}\sum\limits_{t=1}^{T}\theta_{i(u)}},~~\hat{\gamma}^{(m+1)}_{u} = \frac{1}{N}\sum\limits_{i=1}^{N}\theta_{i(u)}, \\
\hat{\Sigma}^{(m+1)}_{p,q|c} &= \frac{\sum\limits_{i=1}^{N}\sum\limits_{t=1}^{T}\theta_{i(c_t)}
    \left\{(\hat{Y}_{ipt}^{(m)}-\hat{\mu}^{(m)}_{p|c})(\hat{Y}_{iqt}^{(m)}-\hat{\mu}^{(m)}_{q|c}) +  \hat{C}_{i(p,q,c_t,t)}^{(m)}\right\}}{\sum\limits_{i=1}^{N}\sum\limits_{t=1}^{T}\theta_{i(c_t)}}.
\end{aligned}
\end{equation}

\noindent Note that $\hat{\gamma}$ above does not include any information from covariates $\textbf{x}_i$. When we are interested in the association between latent trajectory membership and a time-independent covariate, we can instead specify $\hat{\gamma}$ as $\hat{\gamma}(\textbf{x}_i)$, where

\begin{equation}
    \gamma(\textbf{x}_i)= \frac{\exp(\textbf{x}_i\beta_u)}{\sum\limits_{s=1}^S\text{exp}(\textbf{x}_i\beta_s)}, \beta_u \in \mathbb{R}^{P_3}, \beta_1=0,u=1,\dots,S.
\end{equation}
 
Subsequently, we can use the Newton-Raphson in the M-step to update $\beta^{(m+1)}$ where

\begin{equation}
\begin{aligned}
S(\beta^{(m)}) &= \frac{\partial \log L}{\partial \beta_{q u}} = 
\sum_{i=1}^{N}x_{iq}\big(\theta_{i(u)} - \gamma_{u}(\mathbf{x}_{i})\big), \\
H(\beta^{(m)}) &= \frac{\partial^2 \log L}{\partial \beta_{q u}\partial \beta_{q' u'}} 
= \sum_{i=1}^{N}x_{iq}x_{iq'}\!\left[\theta_{i(u)}(\zeta_{uu'}-\theta_{i(u')}) 
- \gamma_{u}(\mathbf{x}_i)(\zeta_{uu'}-\gamma_{u'}(\mathbf{x}_i))\right], \\
\beta^{(m+1)} &= \beta^{(m)} - H(\beta^{(m)})^{-1}S(\beta^{(m)}).
\end{aligned}
\end{equation}
and $\zeta_{ww'}$ is defined as 1 if $u=u'$ and 0 if $u\neq u'$.

\subsection{The second derivative of the MixedLCPM}\label{app2}

This section provides details of the first and second derivative matrix with respect to $[{\boldsymbol\pi}, {\boldsymbol\eta}, {\boldsymbol\mu}, {\boldsymbol\gamma}]$. Eq. (\ref{fst}) shows the elements of first derivatives with respect to parameters $[{\boldsymbol\pi}, {\boldsymbol\eta}, {\boldsymbol\mu}, {\boldsymbol\gamma}]$, and ${\boldsymbol\beta}$. 
\begin{align}\label{fst}
\frac{\partial logL}{\partial\pi_{mk\mid c_t}^{(t)}} &= \sum\limits_{i=1}^{N}\frac{\theta_{i(c_t)}I(Z_{imt}=k)}{\pi_{mk\mid c_t}^{(t)}},~~~~\frac{\partial logL}{\partial\eta_{c_t\mid u}} = \sum\limits_{i=1}^{N}\frac{\theta_{i(u, c_t)}}{\eta_{c_t\mid u}^{(t)}}, \nonumber \\
\frac{\partial logL}{\partial\gamma_{u}} &= \sum\limits_{i=1}^{N}\frac{\theta_{i(u)}}{\gamma_{u}},~~~~
\frac{\partial logL}{\partial\mu_{p\mid c_t}^{(t)}} = \sum\limits_{i=1}^{N}\frac{\theta_{i(c_t)}(Y_{ipt} - \mu_{p\mid c_t}^{(t)})}{\sigma_{p\mid c_t}^2}, \nonumber \\
\frac{\partial logL}{\partial\beta_{q\mid u}} &= \sum\limits_{i=1}^{N}x_{iq}(\theta_{i(u)} - \gamma_{u}({\bf x}_{i}))
\end{align}

\noindent Here, we have $k=1,\ldots,r_m$,~~$p=1,\ldots,P_1$,~~$c_t=1\ldots K$,~~$t=1,\ldots,T$,~~$u=1,\ldots,S$, and $m=1,\ldots,P_2$. Next, the second derivatives with respect to parameters ${\boldsymbol\pi}$ and the others are given as follows:
\begin{align}\label{rho_2nd}
\frac{\partial^2logL}{\partial\pi_{mk\mid c_t}^{(t)}\partial\pi_{m^{'}k^{'}\mid c^{'}_{t^{'}}}^{(t')}} &= \sum\limits_{i=1}^{N}\frac{\left\{\theta_{i(c_t,c'_{t^{'}})}(1 - \zeta_{tt'}) + \theta_{i(c_t)}\zeta_{tt'}(1 - \zeta_{c_tc^{'}_{t^{'}}}) + (1 - \zeta_{mm^{'}}) - \theta_{i(c^{'}_{t^{'}})} \right\}\zeta_{Z_{im^{'}}k^{'}}\zeta_{Z_{im}k}}{\pi_{mk\mid c_t}~\pi_{m^{'}k^{'}\mid c_{t^{'}}}}, \nonumber\\
\frac{\partial^2\log L}{\partial\mu_{p\mid c_t}^{(t)}\partial\pi_{mk\mid c^{'}_{t^{'}}}^{(t^{'})}} &= \sum\limits_{i=1}^{N}\frac{\left\{\theta_{i(c_t,c'_{t^{'}})}(1 - \zeta_{tt'}) + \theta_{i(c_t)}(\zeta_{c_tc^{'}_{t^{'}}}-\theta_{i(c^{'}_{t^{'}})})\right\}(Y_{ipt}-\mu_{p\mid c_t}^{(t)})\zeta_{Z_{imt^{'}}k}}{\sigma_{p\mid c_t}^2},\nonumber \\
\frac{\partial^2logL}{\partial\eta_{c_t\mid u}^{(t)}\partial\pi_{m^{'}k^{'}\mid c^{'}_{t^{'}}}^{(t')}} &= \sum\limits_{i=1}^{N}\frac{\left\{\theta_{i(u,c_t,c^{'}_{t^{'}})}(1 - \zeta_{tt^{'}}) + \theta_{i(u,c_t)}\zeta_{c_tc^{'}_{t^{'}}} - \theta_{i(u^{'}, c^{'}_{t^{'}})}\right\}\zeta_{Z_{im^{'}}k^{'}}}{\eta_{c_t\mid u}^{(t)}\pi_{m^{'}k^{'}\mid c^{'}_{t^{'}}}^{(t')}}, \nonumber\\
\frac{\partial^2logL}{\partial\pi_{mk\mid c_t}^{(t)}\partial\gamma_{u}} &= \sum\limits_{i=1}^{N}\frac{\left\{\theta_{i(u,c_t)}- \theta_{i(c_t)}\theta_{i(u)}\right\}\zeta_{Z_{im}k}}{\pi_{mk\mid c_t}^{(t)}\gamma_{u}}.
\end{align}

\noindent Next, the second derivatives with respect to parameters ${\boldsymbol\mu}$ and the others are given as follows:
\begin{align}\label{mu_2nd}
\frac{\partial^2\log L}{\partial\mu_{p\mid c_t}^{(t)}\partial\mu_{p^{'}\mid c^{'}_{t^{'}}}^{(t^{'})}} &= \sum\limits_{i=1}^{N}\frac{\left\{\theta_{i(c_t,c'_{t^{'}})}(1 - \zeta_{tt'}) + \theta_{i(c_t)}(\zeta_{c_tc^{'}_{t^{'}}}-\theta_{i(c^{'}_{t^{'}})})\right\}(Y_{ipt}-\mu_{p\mid c_t}^{(t)})(Y_{ip^{'}t^{'}}-\mu_{p^{'}\mid c^{'}_{t^{'}}}^{(t^{'})})}{\sigma_{p\mid c_t}^2\sigma_{p^{'}\mid c^{'}_{t^{'}}}^2},\nonumber \\
\frac{\partial^2\log L}{\partial\mu_{p\mid c_t}^{(t)}\partial\eta_{c^{'}_{t^{'}}\mid u}^{(t^{'})}} &= \sum\limits_{i=1}^{N}\frac{\left\{\theta_{i(u,c_t,c'_{t^{'}})}(1 - \zeta_{tt'}) + \theta_{i(u,c_t)}(\zeta_{c_tc^{'}_{t^{'}}}-\theta_{i(c^{'}_{t^{'}})})\right\}(Y_{ipt}-\mu_{p\mid c_t}^{(t)})}{\sigma_{p\mid c_t}^2\eta_{c^{'}_{t^{'}}\mid u}^{(t^{'})}},\nonumber \\
\frac{\partial^2\log L}{\partial\mu_{p\mid c_t}^{(t)}\partial\gamma_{u}} &= \sum\limits_{i=1}^{N}\frac{(\theta_{i(u,c_t)} - \theta_{i(u)}\theta_{i(c_t)})(Y_{ipt}-\mu_{p\mid c_t}^{(t)})}{\sigma_{p\mid c_t}^2\gamma_{u}}.
\end{align}

\noindent Next, the second derivatives with respect to parameters ${\boldsymbol\eta}$ and the others are given as follows:
\begin{align}\label{eta_2nd}
\frac{\partial^2logL}{\partial\eta_{c_t\mid u}^{(t)}\partial\eta_{c^{'}_{t^{'}}\mid u}^{(t')}} &= -\sum\limits_{i=1}^{N}\frac{\theta_{i(u,c_t)}\theta_{i(u^{'},c^{'}_{t^{'}})}}{\eta_{c_t\mid u}^{(t)}\eta_{c^{'}_{t^{'}}\mid u}^{(t')}}, \nonumber\\
\frac{\partial^2logL}{\partial\eta_{c_t\mid u}^{(t)}\partial\gamma_{u^{'}}} &= \sum\limits_{i=1}^{N}\frac{\theta_{i(u,c_t)}(\zeta_{uu^{'}} - \theta_{i(u^{'})})}{\eta_{c_t\mid u}^{(t)}\gamma_{u^{'}}}.
\end{align}

\noindent Finally, the second derivative with respect to parameters ${\boldsymbol\gamma}$ is given as follows:
\begin{align}\label{gamma_2nd}
\frac{\partial^2logL}{\partial\gamma_{u}\partial\gamma_{u^{'}}} &= - \sum\limits_{i=1}^{N}\frac{\theta_{i(u)}\theta_{i(u^{'})}}{\gamma_{u}\gamma_{u^{'}}}.
\end{align}

\noindent From Eq. (\ref{rho_2nd}) to (\ref{gamma_2nd}), subscript ranges are $u,u^{'} = 1,\ldots, S$, $c_t=1,\ldots,K$, $t=1\ldots T$, $k=1,\ldots,r_m$, $m=1,\ldots,P_2$, $p=1,\ldots,P_1$, $u=1\ldots S$, and $\zeta_{ww'}$ is defined as 1 if $w=w'$ and 0 if $w\neq w'$. When subject-level covariates are included in the model, the proportional parameters $\boldsymbol{\gamma}$ become functions of $\boldsymbol{\beta}$. As a result, the second derivatives of $\boldsymbol{\gamma}$ replaced by the second derivatives of $\boldsymbol{\beta}$ as follows:

\begin{align}\label{beta_2nd}
\frac{\partial^2logL}{\partial\pi_{mk\mid c_t}^{(t)}\partial\beta_{qu}} &= \sum\limits_{i=1}^{N}\frac{x_{iq}\left\{\theta_{i(u,c_t)}-\theta_{i(u)}\theta_{i(c_t)})\zeta_{Y_{imt}k}\right\}}{\pi_{mk\mid c_t}^{(t)}}, \nonumber \\
\frac{\partial^2\log L}{\partial\beta_{qu}\partial\beta_{q^{'}u^{'}}} &= \sum\limits_{i=1}^{N}x_{iq}x_{iq^{'}}\left\{\theta_{i(u)}(\zeta_{uu^{'}}-\theta_{i(u^{'})}) - \gamma_{u}({\bf x}_i)(\zeta_{uu^{'}}-\gamma_{u^{'}}({\bf x}_i))\right\}, \nonumber\\
\frac{\partial logL}{\partial\mu_{p\mid c_t}^{(t)}\partial\beta_{q\mid u}} &= 
\sum\limits_{i=1}^{N}\frac{x_{iq}\left\{\theta_{i(u,c_t)} - \theta_{i(u)}\theta_{i(c_t)}\right\}(Y_{ipt}-\mu_{p\mid c_t}^{(t)})}{\sigma_{p\mid c_t}^2}.\nonumber \\
\frac{\partial^2logL}{\partial\eta_{c_t\mid u}^{(t)}\partial\beta_{qu^{'}}} &= \sum\limits_{i=1}^{N}\frac{x_{iq}\left\{\theta_{i(u,c_t)}(\zeta_{uu^{'}}-\theta_{i(u^{'})})\right\}}{\eta_{c_t\mid u}^{(t)}},
\end{align}
where subscript ranges are $q, q^{'} = 1,\ldots,Q$, and $Q$ is the number of covariates. The rest of the subscripts have the same ranges as previously defined. 

\section{Full Simulation Results}

This section presents full simulation results implemented in Section 3. In each scenario, we simulated $900$ data sets using the true parameters, fit the Mixed-LCPM, and then calculate the point and interval estimates. We evaluated model performance using standardized bias (Sbias), root mean-squared error (RMSE), and coverage probability (CP) of the 95\% confidence intervals. Tables \ref{scenario1} through \ref{scenario4nocov} present the results for three sample sizes (i.e., $N=250$, $500$, and $1,000$) across eight scenarios that are different in (i) class separations, (ii) proportions of latent profiles, and (iii) inclusion of covariates. In general, the estimator demonstrated robust performance. For $N=1,000$, coverage probabilities consistently remained near the nominal 0.95 level, and standardized bias remained minimal ($|\text{Sbias}| < 0.1$).  We observed that coverage probabilities were sensitive to sample size. At $N=250$, coverage for certain parameters dropped below 0.90. However, this is expected behavior in mixture modeling due to the asymptotic nature of the standard errors. As sample size increased to $N=500$ and $N=1,000$, coverage improved and RMSE decreased, confirming the consistency of the estimator. Further, for brevity, we only present results on parameters of interest, but we have evaluated performance on all parameters. 

\begin{table}[ht]
\centering
\caption{Coverage, RMSE, and Sbias for Scenario 1 with a covariate across sample sizes}
\label{scenario1}
\begin{tabular}{l|ccc|ccc|ccc}
\hline
Param & \multicolumn{3}{c|}{$N=250$} & \multicolumn{3}{c|}{$N=500$} & \multicolumn{3}{c}{$N=1000$} \\
      & Coverage & RMSE & Sbias & Coverage & RMSE & Sbias & Coverage & RMSE & Sbias \\
\hline
$\gamma_1(\textbf{x}_i)$ & 0.910 & 2.012 & -0.025 & 0.967 & 2.035 & -0.121 & 0.947 & 2.004 & 0.041 \\
$\eta_{1_1|1}$ & 0.953 & 1.571 & 0.050 & 0.957 & 1.563 & 0.130 & 0.943 & 1.551 & -0.017 \\
$\eta_{2_1|1}$ & 0.917 & 1.493 & 0.011 & 0.950 & 1.489 & 0.032 & 0.953 & 1.492 & 0.107 \\
$\eta_{3_1|1}$ & 0.903 & 1.441 & 0.032 & 0.950 & 1.441 & -0.020 & 0.953 & 1.443 & -0.030 \\
$\eta_{1_2|1}$ & 0.903 & 1.445 & -0.031 & 0.927 & 1.440 & 0.007 & 0.950 & 1.443 & -0.127 \\
$\eta_{2_2|1}$ & 0.937 & 1.479 & -0.078 & 0.947 & 1.475 & 0.100 & 0.953 & 1.475 & -0.024 \\
$\eta_{3_2|1}$ & 0.910 & 1.513 & -0.021 & 0.950 & 1.511 & 0.020 & 0.950 & 1.511 & -0.047 \\
$\eta_{1_3|1}$ & 0.937 & 1.432 & -0.006 & 0.957 & 1.430 & -0.078 & 0.937 & 1.425 & 0.018 \\
$\eta_{2_3|1}$ & 0.883 & 1.445 & -0.086 & 0.960 & 1.441 & -0.013 & 0.947 & 1.442 & -0.010 \\
$\eta_{3_3|1}$ & 0.917 & 1.479 & -0.054 & 0.947 & 1.475 & 0.033 & 0.950 & 1.476 & 0.091 \\
$\eta_{1_1|2}$ & 0.907 & 1.513 & -0.049 & 0.957 & 1.510 & -0.047 & 0.940 & 1.511 & -0.085 \\
$\eta_{2_1|2}$ & 0.907 & 1.448 & -0.059 & 0.943 & 1.445 & 0.048 & 0.927 & 1.444 & -0.028 \\
$\eta_{3_1|2}$ & 0.920 & 1.439 & 0.006 & 0.937 & 1.440 & 0.014 & 0.973 & 1.442 & 0.101 \\
$\eta_{1_2|2}$ & 0.950 & 1.469 & 0.098 & 0.957 & 1.462 & -0.068 & 0.923 & 1.468 & -0.005 \\
$\eta_{2_2|2}$ & 0.930 & 1.512 & 0.005 & 0.937 & 1.510 & 0.036 & 0.937 & 1.510 & -0.025 \\
$\eta_{3_2|2}$ & 0.920 & 1.428 & -0.025 & 0.940 & 1.430 & 0.074 & 0.940 & 1.432 & 0.141 \\
$\eta_{1_3|2}$ & 0.913 & 1.444 & 0.001 & 0.943 & 1.444 & -0.011 & 0.940 & 1.440 & -0.054 \\
$\eta_{2_3|2}$ & 0.953 & 1.471 & -0.042 & 0.937 & 1.466 & -0.046 & 0.933 & 1.465 & -0.076 \\
$\eta_{3_3|2}$ & 0.910 & 1.512 & 0.041 & 0.953 & 1.509 & 0.020 & 0.950 & 1.510 & 0.147 \\
\hline
\end{tabular}
\end{table}

\begin{table}[ht]
\centering
\caption{Coverage, RMSE, and Sbias for Scenario 2 with a covariate across sample sizes}
\begin{tabular}{l|ccc|ccc|ccc}
\hline
Param & \multicolumn{3}{c|}{$N=250$} & \multicolumn{3}{c|}{$N=500$} & \multicolumn{3}{c}{$N=1000$} \\
      & Coverage & RMSE & Sbias & Coverage & RMSE & Sbias & Coverage & RMSE & Sbias \\
\hline
$\gamma_1(\textbf{x}_i)$ & 0.887 & 2.299 & -0.102 & 0.913 & 2.075 & -0.065 & 0.930 & 2.046 & -0.039 \\
$\eta_{1_1|1}$ & 0.920 & 3.172 & 0.099 & 0.937 & 1.602 & 0.150 & 0.953 & 1.543 & 0.035 \\
$\eta_{2_1|1}$ & 0.910 & 1.493 & 0.004 & 0.927 & 1.483 & 0.060 & 0.957 & 1.483 & 0.062 \\
$\eta_{3_1|1}$ & 0.903 & 1.433 & -0.074 & 0.927 & 1.438 & -0.097 & 0.933 & 1.438 & -0.012 \\
$\eta_{1_2|1}$ & 0.920 & 1.425 & -0.071 & 0.900 & 1.418 & -0.064 & 0.950 & 1.416 & 0.001 \\
$\eta_{2_2|1}$ & 0.933 & 1.444 & 0.108 & 0.950 & 1.439 & 0.096 & 0.943 & 1.440 & 0.066 \\
$\eta_{3_2|1}$ & 0.920 & 1.483 & 0.027 & 0.953 & 1.484 & 0.125 & 0.953 & 1.485 & 0.131 \\
$\eta_{1_3|1}$ & 0.950 & 1.438 & -0.047 & 0.937 & 1.437 & -0.088 & 0.937 & 1.436 & -0.142 \\
$\eta_{2_3|1}$ & 0.917 & 1.422 & -0.081 & 0.923 & 1.418 & -0.033 & 0.953 & 1.414 & -0.101 \\
$\eta_{3_3|1}$ & 0.930 & 1.436 & 0.042 & 0.940 & 1.439 & 0.001 & 0.940 & 1.441 & 0.042 \\
$\eta_{1_1|2}$ & 0.923 & 1.479 & 0.132 & 0.933 & 1.484 & 0.099 & 0.937 & 1.483 & 0.057 \\
$\eta_{2_1|2}$ & 0.927 & 1.437 & -0.040 & 0.920 & 1.436 & -0.049 & 0.930 & 1.435 & -0.027 \\
$\eta_{3_1|2}$ & 0.927 & 1.416 & 0.046 & 0.910 & 1.413 & 0.031 & 0.957 & 1.415 & 0.041 \\
$\eta_{1_2|2}$ & 0.910 & 1.441 & 0.060 & 0.913 & 1.440 & 0.029 & 0.907 & 1.434 & 0.011 \\
$\eta_{2_2|2}$ & 0.923 & 1.488 & -0.094 & 0.907 & 1.487 & -0.015 & 0.920 & 1.487 & -0.065 \\
$\eta_{3_2|2}$ & 0.960 & 1.436 & -0.065 & 0.910 & 1.435 & -0.135 & 0.937 & 1.437 & -0.113 \\
$\eta_{1_3|2}$ & 0.917 & 1.420 & -0.050 & 0.913 & 1.416 & 0.081 & 0.947 & 1.416 & 0.029 \\
$\eta_{2_3|2}$ & 0.923 & 1.437 & 0.103 & 0.930 & 1.443 & 0.051 & 0.953 & 1.438 & 0.116 \\
$\eta_{3_3|2}$ & 0.900 & 1.484 & -0.035 & 0.917 & 1.483 & 0.048 & 0.927 & 1.484 & -0.073 \\
\hline
\end{tabular}
\end{table}

\begin{table}[ht]
\centering
\caption{Coverage, RMSE, and Sbias for Scenario 3 with a covariate across sample sizes}
\begin{tabular}{l|ccc|ccc|ccc}
\hline
\label{scenario3}
Param & \multicolumn{3}{c|}{$N=250$} & \multicolumn{3}{c|}{$N=500$} & \multicolumn{3}{c}{$N=1000$} \\
      & Coverage & RMSE & Sbias & Coverage & RMSE & Sbias & Coverage & RMSE & Sbias \\
\hline
$\gamma_1(\textbf{x}_i)$ & 0.973 & 2.051 & -0.032 & 0.947 & 2.084 & -0.129 & 0.960 & 2.018 & 0.048 \\
$\eta_{1_1|1}$ & 0.970 & 2.153 & 0.081 & 0.943 & 2.194 & 0.240 & 0.963 & 2.139 & 0.027 \\
$\eta_{2_1|1}$ & 0.953 & 1.504 & 0.021 & 0.927 & 1.499 & 0.084 & 0.937 & 1.502 & 0.098 \\
$\eta_{3_1|1}$ & 0.910 & 1.458 & -0.020 & 0.917 & 1.464 & -0.017 & 0.940 & 1.461 & 0.006 \\
$\eta_{1_2|1}$ & 0.907 & 1.467 & 0.004 & 0.930 & 1.461 & -0.012 & 0.927 & 1.467 & -0.149 \\
$\eta_{2_2|1}$ & 0.920 & 1.498 & -0.022 & 0.923 & 1.495 & 0.088 & 0.947 & 1.495 & -0.010 \\
$\eta_{3_2|1}$ & 0.917 & 1.535 & -0.084 & 0.937 & 1.529 & -0.060 & 0.937 & 1.531 & -0.055 \\
$\eta_{1_3|1}$ & 0.940 & 1.440 & 0.045 & 0.920 & 1.439 & -0.021 & 0.943 & 1.438 & 0.032 \\
$\eta_{2_3|1}$ & 0.927 & 1.467 & -0.107 & 0.940 & 1.464 & 0.070 & 0.947 & 1.463 & -0.064 \\
$\eta_{3_3|1}$ & 0.923 & 1.496 & 0.016 & 0.890 & 1.496 & -0.037 & 0.947 & 1.495 & 0.048 \\
$\eta_{1_1|2}$ & 0.927 & 1.530 & 0.061 & 0.937 & 1.529 & -0.043 & 0.947 & 1.531 & 0.015 \\
$\eta_{2_1|2}$ & 0.947 & 1.463 & -0.046 & 0.953 & 1.462 & 0.002 & 0.943 & 1.462 & 0.036 \\
$\eta_{3_1|2}$ & 0.940 & 1.465 & 0.088 & 0.953 & 1.465 & 0.036 & 0.933 & 1.463 & 0.116 \\
$\eta_{1_2|2}$ & 0.963 & 1.478 & 0.049 & 0.937 & 1.477 & -0.010 & 0.947 & 1.477 & -0.051 \\
$\eta_{2_2|2}$ & 0.917 & 1.532 & -0.005 & 0.957 & 1.529 & 0.002 & 0.970 & 1.529 & -0.089 \\
$\eta_{3_2|2}$ & 0.943 & 1.437 & 0.002 & 0.970 & 1.437 & 0.008 & 0.953 & 1.438 & 0.100 \\
$\eta_{1_3|2}$ & 0.923 & 1.467 & -0.040 & 0.960 & 1.465 & 0.010 & 0.950 & 1.464 & 0.040 \\
$\eta_{2_3|2}$ & 0.943 & 1.478 & -0.024 & 0.947 & 1.478 & -0.063 & 0.943 & 1.475 & -0.105 \\
$\eta_{3_3|2}$ & 0.950 & 1.529 & 0.086 & 0.950 & 1.529 & 0.050 & 0.960 & 1.528 & 0.083 \\
\hline
\end{tabular}
\end{table}

\begin{table}[ht]
\centering
\caption{Coverage, RMSE, and Sbias for Scenario 4 with a covariate across sample sizes}
\label{scenario4}
\begin{tabular}{l|ccc|ccc|ccc}
\hline
Param & \multicolumn{3}{c|}{$N=250$} & \multicolumn{3}{c|}{$N=500$} & \multicolumn{3}{c}{$N=1000$} \\
      & Coverage & RMSE & Sbias & Coverage & RMSE & Sbias & Coverage & RMSE & Sbias \\
\hline
$\gamma_1(\textbf{x}_i)$ & 0.933 & 2.607 & -0.221 & 0.947 & 2.127 & -0.123 & 0.910 & 2.043 & -0.030 \\
$\eta_{1_1|1}$ & 0.907 & 2.908 & 0.280 & 0.947 & 2.296 & 0.229 & 0.940 & 2.227 & 0.144 \\
$\eta_{2_1|1}$ & 0.923 & 1.500 & 0.099 & 0.917 & 1.503 & 0.068 & 0.967 & 1.500 & 0.120 \\
$\eta_{3_1|1}$ & 0.923 & 1.456 & -0.067 & 0.937 & 1.455 & 0.032 & 0.933 & 1.454 & -0.026 \\
$\eta_{1_2|1}$ & 0.900 & 1.429 & 0.000 & 0.943 & 1.439 & -0.111 & 0.957 & 1.438 & -0.139 \\
$\eta_{2_2|1}$ & 0.917 & 1.458 & 0.001 & 0.943 & 1.456 & 0.054 & 0.943 & 1.449 & 0.076 \\
$\eta_{3_2|1}$ & 0.880 & 1.499 & 0.003 & 0.953 & 1.502 & 0.011 & 0.947 & 1.502 & 0.004 \\
$\eta_{1_3|1}$ & 0.920 & 1.455 & -0.009 & 0.913 & 1.455 & -0.014 & 0.933 & 1.457 & -0.099 \\
$\eta_{2_3|1}$ & 0.897 & 1.439 & -0.082 & 0.910 & 1.439 & -0.056 & 0.953 & 1.438 & -0.144 \\
$\eta_{3_3|1}$ & 0.903 & 1.452 & 0.012 & 0.940 & 1.443 & 0.048 & 0.943 & 1.451 & 0.052 \\
$\eta_{1_1|2}$ & 0.907 & 1.505 & 0.035 & 0.943 & 1.499 & -0.024 & 0.947 & 1.502 & 0.147 \\
$\eta_{2_1|2}$ & 0.927 & 1.453 & -0.016 & 0.957 & 1.455 & 0.018 & 0.950 & 1.455 & -0.097 \\
$\eta_{3_1|2}$ & 0.933 & 1.434 & 0.032 & 0.933 & 1.440 & -0.084 & 0.933 & 1.436 & 0.098 \\
$\eta_{1_2|2}$ & 0.910 & 1.447 & -0.025 & 0.957 & 1.454 & 0.041 & 0.933 & 1.453 & -0.018 \\
$\eta_{2_2|2}$ & 0.883 & 1.504 & 0.032 & 0.933 & 1.499 & 0.001 & 0.940 & 1.502 & 0.074 \\
$\eta_{3_2|2}$ & 0.943 & 1.453 & 0.009 & 0.910 & 1.457 & -0.069 & 0.927 & 1.455 & -0.081 \\
$\eta_{1_3|2}$ & 0.917 & 1.433 & -0.035 & 0.930 & 1.437 & 0.047 & 0.947 & 1.435 & -0.022 \\
$\eta_{2_3|2}$ & 0.913 & 1.452 & -0.009 & 0.950 & 1.451 & 0.081 & 0.947 & 1.452 & 0.026 \\
$\eta_{3_3|2}$ & 0.900 & 1.511 & 0.074 & 0.950 & 1.503 & 0.058 & 0.930 & 1.503 & 0.009 \\
\hline
\end{tabular}
\end{table}

\begin{table}[ht]
\centering
\caption{Coverage, RMSE, and Sbias for Scenario 1 with no covariate across sample sizes}
\label{scenario1nocov}
\begin{tabular}{c|ccc|ccc|ccc}
\hline
Param & \multicolumn{3}{c|}{$N=250$} & \multicolumn{3}{c|}{$N=500$} & \multicolumn{3}{c}{$N=1000$} \\
      & Coverage & RMSE & Sbias & Coverage & RMSE & Sbias & Coverage & RMSE & Sbias \\
\hline
$\gamma_{1}$ & 0.923 & 1.387 & -0.075 & 0.937 & 1.389 & 0.013 & 0.950 & 1.389 & 0.009 \\
$\eta_{1_1|1}$ & 0.933 & 1.513 & -0.065 & 0.953 & 1.510 & -0.096 & 0.970 & 1.509 & 0.113 \\
$\eta_{2_1|1}$ & 0.917 & 1.432 & 0.063 & 0.970 & 1.429 & 0.149 & 0.937 & 1.431 & -0.031 \\
$\eta_{3_1|1}$ & 0.950 & 1.461 & 0.040 & 0.930 & 1.463 & 0.029 & 0.957 & 1.465 & 0.017 \\
$\eta_{1_2|1}$ & 0.920 & 1.501 & 0.012 & 0.937 & 1.499 & -0.066 & 0.957 & 1.500 & 0.076 \\
$\eta_{2_2|1}$ & 0.903 & 1.430 & 0.077 & 0.940 & 1.431 & -0.044 & 0.933 & 1.429 & -0.044 \\
$\eta_{3_2|1}$ & 0.923 & 1.507 & -0.042 & 0.940 & 1.509 & 0.063 & 0.950 & 1.510 & 0.013 \\
$\eta_{1_3|1}$ & 0.913 & 1.432 & 0.050 & 0.947 & 1.436 & -0.071 & 0.960 & 1.431 & -0.040 \\
$\eta_{2_3|1}$ & 0.900 & 1.445 & -0.043 & 0.937 & 1.443 & -0.063 & 0.937 & 1.444 & -0.025 \\
$\eta_{3_3|1} $& 0.930 & 1.518 & -0.046 & 0.930 & 1.515 & -0.013 & 0.933 & 1.518 & 0.036 \\
$\eta_{1_1|2} $& 0.933 & 1.432 & -0.161 & 0.927 & 1.431 & 0.063 & 0.950 & 1.431 & -0.050 \\
$\eta_{2_1|2} $& 0.917 & 1.503 & -0.051 & 0.963 & 1.498 & -0.076 & 0.977 & 1.497 & 0.134 \\
$\eta_{3_1|2}$ & 0.920 & 1.455 & 0.107 & 0.920 & 1.445 & -0.006 & 0.953 & 1.446 & -0.025 \\
$\eta_{1_2|2}$ & 0.930 & 1.433 & -0.057 & 0.953 & 1.428 & 0.058 & 0.977 & 1.427 & 0.097 \\
$\eta_{2_2|2}$ & 0.947 & 1.503 & 0.121 & 0.930 & 1.513 & -0.025 & 0.937 & 1.510 & -0.070 \\
$\eta_{3_2|2}$ & 0.920 & 1.434 & -0.061 & 0.957 & 1.430 & 0.035 & 0.947 & 1.431 & -0.013 \\
$\eta_{1_3|2}$ & 0.917 & 1.444 & 0.136 & 0.950 & 1.449 & 0.074 & 0.960 & 1.448 & 0.090 \\
$\eta_{2_3|2}$ & 0.923 & 1.501 & -0.124 & 0.937 & 1.499 & 0.012 & 0.940 & 1.502 & -0.017 \\
$\eta_{3_3|2}$ & 0.923 & 1.427 & 0.075 & 0.907 & 1.427 & -0.042 & 0.957 & 1.429 & -0.098 \\
\hline
\end{tabular}
\end{table}

\begin{table}[ht]
\centering
\caption{Coverage, RMSE, and Sbias for Scenario 2 with no covariate across sample sizes}
\label{scenario2nocov}
\begin{tabular}{c|ccc|ccc|ccc}
\hline
Param & \multicolumn{3}{c|}{$N=250$} & \multicolumn{3}{c|}{$N=500$} & \multicolumn{3}{c}{$N=1000$} \\
      & Coverage & RMSE & Sbias & Coverage & RMSE & Sbias & Coverage & RMSE & Sbias \\
\hline
$\gamma_{1}$ & 0.893 & 1.391 & -0.077 & 0.937 & 1.389 & -0.063 & 0.987 & 1.384 & 0.048 \\
$\eta_{1_1|1}$ & 0.940 & 1.509 & 0.081 & 0.950 & 1.503 & 0.041 & 0.957 & 1.506 & 0.021 \\
$\eta_{2_1|1}$ & 0.980 & 1.421 & -0.057 & 0.943 & 1.426 & -0.071 & 0.960 & 1.425 & -0.031 \\
$\eta_{3_1|1}$ & 0.923 & 1.447 & -0.047 & 0.960 & 1.433 & -0.028 & 0.973 & 1.438 & -0.154 \\
$\eta_{1_2|1}$ & 0.930 & 1.454 & 0.063 & 0.937 & 1.448 & 0.054 & 0.983 & 1.454 & 0.173 \\
$\eta_{2_2|1}$ & 0.947 & 1.405 & 0.045 & 0.960 & 1.401 & 0.008 & 0.947 & 1.401 & 0.024 \\
$\eta_{3_2|1}$ & 0.917 & 1.521 & 0.005 & 0.903 & 1.525 & -0.110 & 0.920 & 1.528 & -0.056 \\
$\eta_{1_3|1}$ & 0.913 & 1.409 & -0.054 & 0.960 & 1.409 & -0.068 & 0.990 & 1.401 & -0.034 \\
$\eta_{2_3|1}$ & 0.917 & 1.401 & 0.064 & 0.927 & 1.396 & 0.081 & 0.987 & 1.396 & 0.137 \\
$\eta_{3_3|1}$ & 0.913 & 1.498 & -0.015 & 0.947 & 1.487 & 0.084 & 0.940 & 1.492 & -0.051 \\
$\eta_{1_1|2}$ & 0.917 & 1.425 & -0.159 & 0.930 & 1.423 & -0.066 & 0.953 & 1.422 & -0.102 \\
$\eta_{2_1|2}$ & 0.920 & 1.472 & 0.107 & 0.920 & 1.472 & 0.138 & 0.943 & 1.470 & 0.037 \\
$\eta_{3_1|2}$ & 0.907 & 1.429 & 0.088 & 0.950 & 1.420 & 0.050 & 0.977 & 1.422 & 0.082 \\
$\eta_{1_2|2}$ & 0.930 & 1.397 & -0.094 & 0.960 & 1.404 & -0.055 & 0.977 & 1.402 & -0.036 \\
$\eta_{2_2|2}$ & 0.937 & 1.526 & -0.037 & 0.943 & 1.526 & -0.094 & 0.943 & 1.530 & 0.104 \\
$\eta_{3_2|2}$ & 0.920 & 1.406 & -0.006 & 0.953 & 1.400 & 0.095 & 0.957 & 1.404 & -0.054 \\
$\eta_{1_3|2}$ & 0.913 & 1.428 & 0.023 & 0.933 & 1.427 & 0.102 & 0.957 & 1.422 & -0.008 \\
$\eta_{2_3|2}$ & 0.920 & 1.477 & -0.059 & 0.957 & 1.476 & -0.101 & 0.977 & 1.471 & -0.023 \\
$\eta_{3_3|2}$ & 0.920 & 1.416 & 0.056 & 0.933 & 1.423 & -0.129 & 0.960 & 1.420 & 0.048 \\
\hline
\end{tabular}
\end{table}

\begin{table}[ht]
\centering
\caption{Coverage, RMSE, and Sbias for Scenario 3 with no covariate across sample sizes}
\label{scenario3nocov}
\begin{tabular}{c|ccc|ccc|ccc}
\hline
Param & \multicolumn{3}{c|}{$N=250$} & \multicolumn{3}{c|}{$N=500$} & \multicolumn{3}{c}{$N=1000$} \\
      & Coverage & RMSE & Sbias & Coverage & RMSE & Sbias & Coverage & RMSE & Sbias \\
\hline
$\gamma_{1}$ & 0.940 & 1.416 & -0.051 & 0.940 & 1.407 & 0.067 & 0.947 & 1.407 & -0.044 \\
$\eta_{1_1|1}$ & 0.953 & 1.516 & -0.017 & 0.950 & 1.511 & -0.006 & 0.963 & 1.511 & 0.091 \\
$\eta_{2_1|1}$ & 0.943 & 1.430 & 0.124 & 0.940 & 1.432 & 0.153 & 0.950 & 1.433 & -0.052 \\
$\eta_{3_1|1}$ & 0.960 & 1.473 & -0.053 & 0.940 & 1.467 & -0.038 & 0.950 & 1.464 & 0.004 \\
$\eta_{1_2|1}$ & 0.943 & 1.502 & 0.052 & 0.920 & 1.504 & 0.063 & 0.943 & 1.501 & 0.051 \\
$\eta_{2_2|1}$ & 0.947 & 1.436 & -0.031 & 0.917 & 1.431 & -0.088 & 0.957 & 1.432 & -0.060 \\
$\eta_{3_2|1}$ & 0.947 & 1.514 & 0.020 & 0.943 & 1.512 & 0.070 & 0.960 & 1.509 & 0.021 \\
$\eta_{1_3|1}$ & 0.937 & 1.433 & 0.090 & 0.927 & 1.438 & -0.175 & 0.940 & 1.432 & 0.032 \\
$\eta_{2_3|1}$ & 0.917 & 1.450 & 0.017 & 0.893 & 1.446 & -0.121 & 0.923 & 1.447 & 0.024 \\
$\eta_{3_3|1}$ & 0.897 & 1.520 & -0.098 & 0.907 & 1.521 & 0.053 & 0.960 & 1.518 & 0.000 \\
$\eta_{1_1|2}$ & 0.923 & 1.434 & -0.062 & 0.953 & 1.430 & 0.073 & 0.947 & 1.432 & -0.031 \\
$\eta_{2_1|2}$ & 0.927 & 1.499 & 0.032 & 0.957 & 1.501 & -0.005 & 0.940 & 1.500 & -0.008 \\
$\eta_{3_1|2}$ & 0.927 & 1.452 & 0.010 & 0.943 & 1.450 & -0.033 & 0.937 & 1.447 & 0.022 \\
$\eta_{1_2|2}$ & 0.930 & 1.431 & -0.095 & 0.940 & 1.430 & 0.089 & 0.947 & 1.429 & 0.026 \\
$\eta_{2_2|2}$ & 0.930 & 1.516 & 0.145 & 0.950 & 1.513 & -0.008 & 0.933 & 1.511 & 0.027 \\
$\eta_{3_2|2}$ & 0.957 & 1.430 & -0.005 & 0.953 & 1.432 & 0.080 & 0.890 & 1.433 & -0.068 \\
$\eta_{1_3|2}$ & 0.920 & 1.453 & 0.027 & 0.960 & 1.449 & 0.024 & 0.963 & 1.447 & 0.058 \\
$\eta_{2_3|2}$ & 0.907 & 1.500 & -0.061 & 0.970 & 1.502 & 0.028 & 0.940 & 1.501 & 0.012 \\
$\eta_{3_3|2}$ & 0.933 & 1.430 & 0.046 & 0.937 & 1.430 & -0.092 & 0.953 & 1.431 & -0.084 \\
\hline
\end{tabular}
\end{table}

\begin{table}[ht]
\centering
\caption{Coverage, RMSE, and Sbias for Scenario 4 with no covariate across sample sizes}
\label{scenario4nocov}
\begin{tabular}{c|ccc|ccc|ccc}
\hline
Param & \multicolumn{3}{c|}{$N=250$} & \multicolumn{3}{c|}{$N=500$} & \multicolumn{3}{c}{$N=1000$} \\
      & Coverage & RMSE & Sbias & Coverage & RMSE & Sbias & Coverage & RMSE & Sbias \\
\hline
$\gamma_{1}$ & 0.930 & 1.407 & -0.308 & 0.940 & 1.401 & -0.165 & 0.977 & 1.406 & -0.075 \\
$\eta_{1_1|1}$ & 0.870 & 1.520 & -0.060 & 0.923 & 1.510 & 0.073 & 0.953 & 1.508 & 0.057 \\
$\eta_{2_1|1}$ & 0.967 & 1.430 & 0.026 & 0.960 & 1.423 & 0.035 & 0.953 & 1.426 & -0.048 \\
$\eta_{3_1|1}$ & 0.887 & 1.458 & 0.073 & 0.940 & 1.437 & -0.115 & 0.967 & 1.443 & -0.108 \\
$\eta_{1_2|1}$ & 0.883 & 1.463 & -0.030 & 0.927 & 1.463 & -0.002 & 0.967 & 1.449 & 0.031 \\
$\eta_{2_2|1}$ & 0.910 & 1.401 & -0.024 & 0.920 & 1.397 & 0.090 & 0.960 & 1.401 & 0.132 \\
$\eta_{3_2|1}$ & 0.940 & 1.528 & 0.073 & 0.943 & 1.537 & -0.105 & 0.947 & 1.528 & -0.093 \\
$\eta_{1_3|1}$ & 0.880 & 1.408 & 0.058 & 0.933 & 1.403 & -0.076 & 0.983 & 1.409 & -0.083 \\
$\eta_{2_3|1}$ & 0.890 & 1.402 & -0.043 & 0.960 & 1.394 & 0.044 & 0.953 & 1.391 & 0.053 \\
$\eta_{3_3|1}$ & 0.877 & 1.497 & -0.114 & 0.907 & 1.494 & 0.019 & 0.940 & 1.494 & 0.071 \\
$\eta_{1_1|2}$ & 0.937 & 1.424 & 0.027 & 0.943 & 1.423 & -0.053 & 0.960 & 1.424 & -0.071 \\
$\eta_{2_1|2}$ & 0.930 & 1.472 & 0.035 & 0.953 & 1.470 & 0.091 & 0.957 & 1.471 & 0.019 \\
$\eta_{3_1|2}$ & 0.923 & 1.422 & 0.111 & 0.953 & 1.426 & 0.160 & 0.977 & 1.422 & 0.077 \\
$\eta_{1_2|2}$ & 0.923 & 1.408 & -0.200 & 0.953 & 1.398 & -0.063 & 0.967 & 1.400 & -0.012 \\
$\eta_{2_2|2}$ & 0.970 & 1.529 & -0.040 & 0.953 & 1.526 & -0.102 & 0.977 & 1.531 & -0.186 \\
$\eta_{3_2|2}$ & 0.940 & 1.404 & 0.060 & 0.957 & 1.406 & 0.067 & 0.963 & 1.403 & -0.008 \\
$\eta_{1_3|2}$ & 0.927 & 1.423 & 0.124 & 0.963 & 1.426 & 0.132 & 0.973 & 1.422 & 0.096 \\
$\eta_{2_3|2}$ & 0.943 & 1.476 & -0.131 & 0.967 & 1.472 & -0.096 & 0.970 & 1.476 & -0.068 \\
$\eta_{3_3|2}$ & 0.947 & 1.422 & -0.056 & 0.950 & 1.423 & -0.104 & 0.970 & 1.421 & -0.113 \\
\hline
\end{tabular}
\end{table}

\end{document}

%% file: structure.tex
\usepackage{amsmath,amsfonts,stmaryrd,amssymb} 

\usepackage{enumerate} 

\usepackage[ruled]{algorithm2e} 

\usepackage[framemethod=tikz]{mdframed} 

\usepackage{listings} 
\usepackage{geometry} 

\usepackage[utf8]{inputenc} 
\usepackage[T1]{fontenc} 
\usepackage{times}
\usepackage{cite}

\mdfdefinestyle{question}{
	innertopmargin=1.2\baselineskip,
	innerbottommargin=0.8\baselineskip,
	roundcorner=5pt,
	nobreak,
	singleextra={%
		\draw(P-|O)node[xshift=1em,anchor=west,fill=white,draw,rounded corners=5pt]{%
		Question \theQuestion\questionTitle};
	},
}

\newcounter{Question} 

\mdfdefinestyle{warning}{
	topline=false, bottomline=false,
	leftline=false, rightline=false,
	nobreak,
	singleextra={%
		\draw(P-|O)++(-0.5em,0)node(tmp1){};
		\draw(P-|O)++(0.5em,0)node(tmp2){};
		\fill[black,rotate around={45:(P-|O)}](tmp1)rectangle(tmp2);
		\node at(P-|O){\color{white}\scriptsize\bf !};
		\draw[very thick](P-|O)++(0,-1em)--(O);
	}
}



%% file: References.bib
@article{bateman_two_2023,
	title = {Two {Phase} 3 {Trials} of {Gantenerumab} in {Early} {Alzheimer}’s {Disease}},
	volume = {389},
	issn = {0028-4793},
	url = {https://www.nejm.org/doi/full/10.1056/NEJMoa2304430},
	doi = {10.1056/NEJMoa2304430},
	number = {20},
	urldate = {2025-12-09},
	journal = {New England Journal of Medicine},
	author = {Bateman, Randall J. and Smith, Janice and Donohue, Michael C. and Delmar, Paul and Abbas, Rachid and Salloway, Stephen and Wojtowicz, Jakub and Blennow, Kaj and Bittner, Tobias and Black, Sandra E. and Klein, Gregory and Boada, Mercè and Grimmer, Timo and Tamaoka, Akira and Perry, Richard J. and Turner, R. Scott and Watson, David and Woodward, Michael and Thanasopoulou, Angeliki and Lane, Christopher and Baudler, Monika and Fox, Nick C. and Cummings, Jeffrey L. and Fontoura, Paulo and Doody, Rachelle S.},
	month = nov,
	year = {2023},
	note = {Publisher: Massachusetts Medical Society
\_eprint: https://www.nejm.org/doi/pdf/10.1056/NEJMoa2304430},
	pages = {1862--1876},
}

@article{koo2020bayesian,
  title={Bayesian nonparametric latent class model for longitudinal data},
  author={Koo, Wonmo and Kim, Heeyoung},
  journal={Statistical Methods in Medical Research},
  volume={29},
  number={11},
  pages={3381--3395},
  year={2020},
  publisher={SAGE Publications Sage UK: London, England}
}

@article{better2023alzheimer,
  title={Alzheimer’s disease facts and figures},
  author={Better, MAPPING A},
  journal={Alzheimers Dement},
  volume={19},
  number={4},
  pages={1598--1695},
  year={2023}
}

@incollection{muthen_chapter_2004,
	address = {2455 Teller Road, Thousand Oaks California 91320 United States of America},
	title = {Chapter 19: {Latent} {Variable} {Analysis}: {Growth} {Mixture} {Modeling} and {Related} {Techniques} for {Longitudinal} {Data}},
	isbn = {978-0-7619-2359-6 978-1-4129-8631-1},
	shorttitle = {Latent {Variable} {Analysis}},
	url = {https://methods.sagepub.com/book/the-sage-handbook-of-quantitative-methodology-for-the-social-sciences/n19.xml},
	urldate = {2025-11-12},
	booktitle = {The {SAGE} {Handbook} of {Quantitative} {Methodology} for the {Social} {Sciences}},
	publisher = {SAGE Publications, Inc.},
	author = {Muthén, Bengt},
	year = {2004},
	pages = {346--369},
}

@article{lee_bayesian_2025,
	title = {Bayesian {Multilevel} {Latent} {Class} {Profile} {Analysis}: {Inference} and {Estimation} for {Exploring} the {Diverse} {Pathways} to {Academic} {Proficiency}},
	volume = {60},
	issn = {0027-3171},
	shorttitle = {Bayesian {Multilevel} {Latent} {Class} {Profile} {Analysis}},
	url = {https://doi.org/10.1080/00273171.2025.2501341},
	doi = {10.1080/00273171.2025.2501341},
	number = {5},
	urldate = {2025-11-20},
	journal = {Multivariate Behavioral Research},
	author = {Lee, JungWun and McCoach, D. Betsy and Harel, Ofer and Chung, Hwan},
	month = sep,
	year = {2025},
	pmid = {40400340},
	note = {Publisher: Routledge
\_eprint: https://doi.org/10.1080/00273171.2025.2501341},
	pages = {954--972},
}

@article{lee_latent_2025,
	title = {Latent class profile model with time-dependent covariates: a study on symptom patterning of patients for head and neck cancer},
	volume = {52},
	issn = {0266-4763},
	shorttitle = {Latent class profile model with time-dependent covariates},
	doi = {10.1080/02664763.2024.2435997},
	language = {eng},
	number = {8},
	journal = {Journal of Applied Statistics},
	author = {Lee, Jung Wun and Yackel, Hayley Dunnack},
	year = {2025},
	pmid = {40497159},
	pmcid = {PMC12147489},
	pages = {1628--1648},
}

@article{wu_convergence_1983,
	title = {On the {Convergence} {Properties} of the {EM} {Algorithm}},
	volume = {11},
	issn = {0090-5364, 2168-8966},
	url = {https://projecteuclid.org/journals/annals-of-statistics/volume-11/issue-1/On-the-Convergence-Properties-of-the-EM-Algorithm/10.1214/aos/1176346060.full},
	doi = {10.1214/aos/1176346060},
	number = {1},
	urldate = {2025-11-12},
	journal = {The Annals of Statistics},
	author = {Wu, C. F. Jeff},
	month = mar,
	year = {1983},
	note = {Publisher: Institute of Mathematical Statistics},
	pages = {95--103},
}

@article{schwarz_estimating_1978,
	title = {Estimating the {Dimension} of a {Model}},
	volume = {6},
	url = {https://ui.adsabs.harvard.edu/abs/1978AnSta...6..461S},
	urldate = {2025-11-12},
	journal = {Annals of Statistics},
	author = {Schwarz, Gideon},
	month = jul,
	year = {1978},
	note = {ADS Bibcode: 1978AnSta...6..461S},
	pages = {461--464},
}

@misc{national_alzheimers_project_act_alzheimers_nodate,
	title = {Alzheimer’s {Disease} {Sequencing} {Project}},
	howpublished = {\url{https://www.nia.nih.gov/research/dn/alzheimers-disease-sequencing-project}},
	language = {en},
	urldate = {2025-11-12},
	journal = {National Institute on Aging},
	author = {{National Alzheimer's Project Act}},
}

@article{goodman_exploratory_1974,
	title = {Exploratory latent structure analysis using both identifiable and unidentifiable models},
	volume = {61},
	issn = {0006-3444},
	url = {https://doi.org/10.1093/biomet/61.2.215},
	doi = {10.1093/biomet/61.2.215},
	number = {2},
	urldate = {2025-11-12},
	journal = {Biometrika},
	author = {Goodman, Leo A.},
	month = aug,
	year = {1974},
	pages = {215--231},
}

@article{muthen_integrating_2000,
	title = {Integrating {Person}-{Centered} and {Variable}-{Centered} {Analyses}: {Growth} {Mixture} {Modeling} {With} {Latent} {Trajectory} {Classes}},
	volume = {24},
	issn = {1530-0277},
	shorttitle = {Integrating {Person}-{Centered} and {Variable}-{Centered} {Analyses}},
	url = {https://onlinelibrary.wiley.com/doi/abs/10.1111/j.1530-0277.2000.tb02070.x},
	doi = {10.1111/j.1530-0277.2000.tb02070.x},
	language = {en},
	number = {6},
	urldate = {2025-11-12},
	journal = {Alcoholism: Clinical and Experimental Research},
	author = {Muthén, Bengt and Muthén, Linda K.},
	year = {2000},
	note = {\_eprint: https://onlinelibrary.wiley.com/doi/pdf/10.1111/j.1530-0277.2000.tb02070.x},
	pages = {882--891},
}

@article{jung_introduction_2008,
	title = {An {Introduction} to {Latent} {Class} {Growth} {Analysis} and {Growth} {Mixture} {Modeling}},
	volume = {2},
	copyright = {© 2007 The Authors},
	issn = {1751-9004},
	url = {https://onlinelibrary.wiley.com/doi/abs/10.1111/j.1751-9004.2007.00054.x},
	doi = {10.1111/j.1751-9004.2007.00054.x},
	language = {en},
	number = {1},
	urldate = {2025-11-12},
	journal = {Social and Personality Psychology Compass},
	author = {Jung, Tony and Wickrama, K. a. S.},
	year = {2008},
	note = {\_eprint: https://compass.onlinelibrary.wiley.com/doi/pdf/10.1111/j.1751-9004.2007.00054.x},
	pages = {302--317},
}

@article{lee_bayesian_2021,
	title = {Bayesian multivariate latent class profile analysis: {Exploring} the developmental progression of youth depression and substance use},
	volume = {161},
	issn = {0167-9473},
	shorttitle = {Bayesian multivariate latent class profile analysis},
	url = {https://www.sciencedirect.com/science/article/pii/S0167947321000955},
	doi = {10.1016/j.csda.2021.107261},
	urldate = {2025-11-12},
	journal = {Computational Statistics \& Data Analysis},
	author = {Lee, Jung Wun and Chung, Hwan and Jeon, Saebom},
	month = sep,
	year = {2021},
	pages = {107261},
}

@book{lazarsfeld_latent_1968,
	title = {Latent {Structure} {Analysis}},
	isbn = {978-0-395-04768-2},
	language = {en},
	publisher = {Houghton, Mifflin},
	author = {Lazarsfeld, Paul F. and Henry, Neil W.},
	year = {1968},
	note = {Google-Books-ID: jsFPAQAAIAAJ},
}

@article{chung_latent_2008,
	title = {Latent transition analysis: inference and estimation},
	volume = {27},
	issn = {0277-6715},
	shorttitle = {Latent transition analysis},
	url = {https://pmc.ncbi.nlm.nih.gov/articles/PMC3158980/},
	doi = {10.1002/sim.3130},
	number = {11},
	urldate = {2025-11-12},
	journal = {Statistics in medicine},
	author = {Chung, Hwan and Lanza, Stephanie T. and Loken, Eric},
	month = may,
	year = {2008},
	pmid = {18069720},
	pmcid = {PMC3158980},
	pages = {1834--1854},
}

@article{lee_latent_2024,
	title = {A latent class selection model for categorical response variables with nonignorably missing data},
	volume = {17},
	doi = {10.4310/22-SII753},
	journal = {Statistics and Its Interface},
	author = {Lee, Jung Wun and Harel, Ofer},
	month = jan,
	year = {2024},
	pages = {635--648},
}

@article{proust-lima_estimation_2017,
	title = {Estimation of {Extended} {Mixed} {Models} {Using} {Latent} {Classes} and {Latent} {Processes}: {The} {R} {Package} lcmm},
	volume = {78},
	copyright = {Copyright (c) 2017 Cécile Proust-Lima, Viviane Philipps, Benoit Liquet},
	issn = {1548-7660},
	shorttitle = {Estimation of {Extended} {Mixed} {Models} {Using} {Latent} {Classes} and {Latent} {Processes}},
	url = {https://doi.org/10.18637/jss.v078.i02},
	doi = {10.18637/jss.v078.i02},
	language = {en},
	urldate = {2025-11-12},
	journal = {Journal of Statistical Software},
	author = {Proust-Lima, Cécile and Philipps, Viviane and Liquet, Benoit},
	month = jun,
	year = {2017},
	pages = {1--56},
}

@article{proust-lima_describing_2023,
	title = {Describing complex disease progression using joint latent class models for multivariate longitudinal markers and clinical endpoints},
	volume = {42},
	issn = {1097-0258},
	doi = {10.1002/sim.9844},
	language = {eng},
	number = {22},
	journal = {Statistics in Medicine},
	author = {Proust-Lima, Cécile and Saulnier, Tiphaine and Philipps, Viviane and Traon, Anne Pavy-Le and Péran, Patrice and Rascol, Olivier and Meissner, Wassilios G. and Foubert-Samier, Alexandra},
	month = sep,
	year = {2023},
	pmid = {37461227},
	pages = {3996--4014},
}

@article{tadde_dynamic_2020,
	title = {Dynamic modeling of multivariate dimensions and their temporal relationships using latent processes: {Application} to {Alzheimer}'s disease},
	volume = {76},
	issn = {1541-0420},
	shorttitle = {Dynamic modeling of multivariate dimensions and their temporal relationships using latent processes},
	doi = {10.1111/biom.13168},
	language = {eng},
	number = {3},
	journal = {Biometrics},
	author = {Taddé, Bachirou O. and Jacqmin-Gadda, Hélène and Dartigues, Jean-François and Commenges, Daniel and Proust-Lima, Cécile},
	month = sep,
	year = {2020},
	pmid = {31647111},
	pages = {886--899},
}

@article{lee_latent_2025-2,
	title = {A latent class pattern mixture model for nonignorable nonresponses in multivariate categorical data},
	volume = {40},
	issn = {1613-9658},
	url = {https://doi.org/10.1007/s00180-025-01627-0},
	doi = {10.1007/s00180-025-01627-0},
	language = {en},
	number = {8},
	urldate = {2025-11-12},
	journal = {Computational Statistics},
	author = {Lee, Jungwun and Lloyd Sieger, Margaret and Phillips, Jon D.},
	month = nov,
	year = {2025},
	pages = {4367--4397},
}

@article{bandeen-roche_latent_1997,
	title = {Latent {Variable} {Regression} for {Multiple} {Discrete} {Outcomes}},
	volume = {92},
	issn = {0162-1459},
	url = {https://doi.org/10.1080/01621459.1997.10473658},
	doi = {10.1080/01621459.1997.10473658},
	number = {440},
	urldate = {2025-09-10},
	journal = {Journal of the American Statistical Association},
	author = {Bandeen-roche, Karen and Miglioretti, Diana L. and Zeger, Scott L. and Rathouz, Paul J.},
	month = dec,
	year = {1997},
	note = {Publisher: ASA Website
\_eprint: https://doi.org/10.1080/01621459.1997.10473658},
	pages = {1375--1386},
}

@article{nguena_nguefack_trajectory_2020,
	title = {Trajectory {Modelling} {Techniques} {Useful} to {Epidemiological} {Research}: {A} {Comparative} {Narrative} {Review} of {Approaches}},
	volume = {12},
	issn = {1179-1349},
	shorttitle = {Trajectory {Modelling} {Techniques} {Useful} to {Epidemiological} {Research}},
	url = {https://www.ncbi.nlm.nih.gov/pmc/articles/PMC7608582/},
	doi = {10.2147/CLEP.S265287},
	urldate = {2025-08-11},
	journal = {Clinical Epidemiology},
	author = {Nguena Nguefack, Hermine Lore and Pagé, M Gabrielle and Katz, Joel and Choinière, Manon and Vanasse, Alain and Dorais, Marc and Samb, Oumar Mallé and Lacasse, Anaïs},
	month = oct,
	year = {2020},
	pmid = {33154677},
	pmcid = {PMC7608582},
	pages = {1205--1222},
}

@article{proust-lima_describing_2023-1,
	title = {Describing complex disease progression using joint latent class models for multivariate longitudinal markers and clinical endpoints},
	volume = {42},
	issn = {1097-0258},
	doi = {10.1002/sim.9844},
	language = {eng},
	number = {22},
	journal = {Statistics in Medicine},
	author = {Proust-Lima, Cécile and Saulnier, Tiphaine and Philipps, Viviane and Traon, Anne Pavy-Le and Péran, Patrice and Rascol, Olivier and Meissner, Wassilios G. and Foubert-Samier, Alexandra},
	month = sep,
	year = {2023},
	pmid = {37461227},
	pages = {3996--4014},
}

@article{bartolucci_overview_2010,
	title = {An overview of latent {Markov} models for longitudinal categorical data},
	url = {http://arxiv.org/abs/1003.2804},
	doi = {10.48550/arXiv.1003.2804},
	urldate = {2025-04-22},
	publisher = {arXiv},
	author = {Bartolucci, F. and Farcomeni, A. and Pennoni, F.},
	month = mar,
	year = {2010},
	note = {arXiv:1003.2804 [math]},
}

@article{dempster_maximum_1977,
	title = {Maximum {Likelihood} from {Incomplete} {Data} via the {EM} {Algorithm}},
	volume = {39},
	issn = {0035-9246},
	url = {https://www.jstor.org/stable/2984875},
	number = {1},
	urldate = {2025-04-22},
	journal = {Journal of the Royal Statistical Society. Series B (Methodological)},
	author = {Dempster, A. P. and Laird, N. M. and Rubin, D. B.},
	year = {1977},
	note = {Publisher: [Royal Statistical Society, Oxford University Press]},
	pages = {1--38},
}

@article{chung_latent_2011,
	title = {Latent class profile analysis: an application to stage-sequential process in early-onset drinking behaviours},
	volume = {174},
	issn = {0964-1998},
	shorttitle = {Latent class profile analysis},
	url = {https://www.ncbi.nlm.nih.gov/pmc/articles/PMC4906792/},
	doi = {10.1111/j.1467-985X.2010.00674.x},
	number = {3},
	urldate = {2025-04-16},
	journal = {Journal of the Royal Statistical Society. Series A, (Statistics in Society)},
	author = {Chung, Hwan and Anthony, James C. and Schafer, Joseph L.},
	month = jul,
	year = {2011},
	pmid = {27313406},
	pmcid = {PMC4906792},
	pages = {689--712},
}

@article{proust-lima_analysis_2013,
	title = {Analysis of multivariate mixed longitudinal data: a flexible latent process approach},
	volume = {66},
	issn = {2044-8317},
	shorttitle = {Analysis of multivariate mixed longitudinal data},
	doi = {10.1111/bmsp.12000},
	language = {eng},
	number = {3},
	journal = {The British Journal of Mathematical and Statistical Psychology},
	author = {Proust-Lima, Cécile and Amieva, Hélène and Jacqmin-Gadda, Hélène},
	month = nov,
	year = {2013},
	pmid = {23082854},
	pages = {470--487},
}
